\documentclass[aps,prb,reprint,groupedaddress, showpacs]{revtex4-2}
\usepackage{graphicx,graphics}
\usepackage{epstopdf}
\usepackage{dcolumn}
\usepackage{amsmath,amssymb,amsfonts}
\usepackage{latexsym,verbatim}
\usepackage{bm}
\usepackage{bbold}
\usepackage{color}
\usepackage{ulem}
\usepackage[breaklinks=false,colorlinks,citecolor=blue,linkcolor=blue,urlcolor=blue]{hyperref}
\usepackage[abs]{overpic}
\usepackage{textcomp}
\usepackage{mathtools}
\usepackage{braket}
\usepackage{soul}

\begin{document}

\title{Theory of multi-dimensional quantum capacitance and its
application to spin and
charge discrimination in quantum-dot arrays}

\author{Andrea Secchi}
\email{andrea.secchi@nano.cnr.it}
\affiliation{Centro S3, CNR-Istituto di Nanoscienze, I-41125 Modena,
Italy}
\author{Filippo Troiani}
\affiliation{Centro S3, CNR-Istituto di Nanoscienze, I-41125 Modena,
Italy}

\begin{abstract}
Quantum states of a few-particle system capacitively coupled to a metal
gate can be discriminated by measuring the quantum capacitance, which
can be identified with the second derivative of the system energy with
respect to the gate voltage. This approach is here generalized to the
multi-voltage case, through the introduction of the quantum capacitance
matrix. The matrix formalism allows us to determine the dependence of
the quantum capacitance on the direction of the voltage oscillations in
the parameter space, and to identify the optimal combination of gate
voltages. As a representative example, this approach is applied to the case of a quantum-dot array,
described in terms of a Hubbard model. Here, we first identify the
potentially relevant regions in the multi-dimensional voltage space with
the boundaries between charge stability regions, determined within a
semiclassical approach. Then, we quantitatively characterize such
boundaries by means of the quantum capacitance matrix. Altogether, this
provides a procedure for optimizing the discrimination between states
with different particle numbers and/or total spins.
\end{abstract}

\date{\today}

\maketitle

\section{Introduction}

The role played by quantum effects in determining the capacitance of a
nanoscopic system has been the subject of a long investigation. These
effects include electron statistics \cite{Luryi88}, band structure and
related density of states \cite{Fang07}, Coulomb interactions
\cite{Kopp2009a,Berthod2021a}, reduced dimensionality \cite{Ilani2006},
presence of charge impurities \cite{Xia2009}, just to mention a few.

Semiconductor quantum dots represent a particular class of nanoscopic
systems, which allows an extreme degree of control on the number of
confined charge carriers, and on their orbital and spin properties
\cite{vanderWiel02a,Hanson07a,Zwanenburg13a}. When coupled to a
classical circuit, the dots affect its properties by contributing a
complex parametric impedance
\cite{Ciccarelli11a,Chorley2012a,Cheong2002a,Persson2010a,Frey2012a}.
In particular, the reactive component of such impedance has a capacitive
character if the carriers' response to the applied voltage is faster
than the probing frequency \cite{Cottet2011a}.
There, a further distinction is typically made between tunneling and
quantum capacitance (QC): the former one results from population
redistribution processes amongst the eigenstates, induced by
nonadiabatic or incoherent
transitions; the latter one, on which we focus hereafter, is related to
adiabatic evolution and to the curvature of the energy levels
\cite{Mizuta2017a}.

Being the QC dependent on the charge and the spin of few-electron
systems in tunnel-coupled quantum dots, its measurement can been
exploited to discriminate between quantum states \cite{Vigneau22a}. This
mechanism, at sufficiently high values of the signal-to-noise ratio and
of the bandwidth, has in fact enabled the dispersive readout of
different spin qubits in semiconductor quantum dots
\cite{Barthel2009a,Crippa2019a,Urdampilleta2019a,West2019a,Pakkiam2018a,Zheng2019a}.
A particularly promising approach in terms of scalability is represented
by gate reflectometry, where the readout process is based on the quantum
contribution of the qubit to the metal gate capacitance, and on the
resulting frequency shift of a connected resonator
\cite{Colless13a,Gonzalez-Zalba2015a}.

Essentially, the QC is large at working points where small changes in
the gate voltage induce large and quantum-state dependent fluctuations
in the occupation of the underlying dot(s). This criterion can be
directly derived from the identification of the QC with the derivative
of the dot occupation with respect to the voltage $V$
\cite{Duty2005a,John04,Shevchenko2015,Crippa2017a}. Alternatively, convenient working points can be identified with those voltages at which the energy levels
undergo avoided crossings. This is consistent with the definition of the
QC in terms of the second derivative of the relevant energy eigenvalue
with respect to $V$ \cite{Gonzalez-Zalba2016a,Petersson2010a}. In double
quantum dots, single- and two-particle systems display an avoided level
crossing as a function of the detuning voltage, as the ground state
undergoes a transition from one charge configuration to the other, which
leads to a convergence of the two above criteria. In fact, within a
qubit model, the definitions of the QC based on the first derivative of
a dot occupation and on the second derivative of the energy eigenstate
can be formally shown to coincide \cite{Mizuta2017a}. However, the range
of validity of such correspondence still needs to be determined.

In the following, we consider a more general case, where the energy of a confined
few-particle system depends on $M>1$ tunable gate voltages. The usual concept of a
scalar QC must thus be replaced with that of the {\it quantum
capacitance matrix}, whose elements are related to the second mixed
derivatives of the system energy with respect to voltages. 
This matrix differs from the ``classical'' capacitance matrix, which applies to systems of ideal conductors and provides linear relations between their charges and voltages \cite{vanderWiel02a}.
While the scalar QC of a system is probed by applying a small
oscillating voltage to the metal gate, in the multidimensional approach
the measured QC depends on the combination of the perturbations coherently applied
to $M$ gates \cite{Mills2019}, corresponding to a specific direction in the
$M$-dimensional voltage space. Such dependence is fully captured by the
QC matrix, which thus allows the identification of an {\it optimal
direction} in the voltage space, along which the measured QC reaches its
maximum absolute value for a given working point. The optimal direction coincides with that of the eigenvector of the QC matrix whose eigenvalue has the largest modulus.
Besides, the functional dependence of the QC matrix on the charge
density and its derivatives with respect to the voltages --- derived
through the Hellmann-Feynman theorem --- generalizes the above mentioned
relation derived for the case of an effective two-level system.

Applying the approach to a Hubbard model, which provides a simple
representation of an array of tunnel-coupled quantum dots, we then derive a discretized
version of the general equations. Here, the QC matrix is related via the
Hellmann-Feynman theorem to the dot occupations, rather than the charge
density characterizing a continuous system. The $M$-dimensional voltage
space can be partitioned in charge stability regions (CSRs), inside
which the fluctuations in the ground-state dot occupations are
suppressed. The relevant features of the QC are shown to occur at the
boundaries between adjacent CSRs; these features include peaks and
plateaus -- the latter being a peculiarity of the multi-gate system.
While the positions of the boundaries are identified on the basis of a
semiclassical approach, the behavior of the QC matrix can only be
determined by means of a fully quantum approach, as the QC features
depend crucially on the quantum-tunneling processes connecting the CSRs.
In fact, the quantum approach provides the optimal directions and the
corresponding values of the QC. Interestingly, the dependence of the
charge stability pattern on the number of particles ($N$) and on the
total spin ($S$) allows the identification of regions in the voltage
space that are suitable for the discrimination between different values
of $N$ and/or $S$, thus extending the approach that has been implemented
for the readout of spin qubits in quantum dots.

The remainder of the paper is organized as follows. In Sec.~\ref{sec:
MQC} we derive general expressions for the QC in the one-dimensional
(Subsec.~\ref{subsec:SQC}) and in the multidimensional
(Subsec.~\ref{subsec:VQC}) case. Section \ref{sec: Hubbard} is devoted
to the application of the approach to a quantum-dot array, represented
in terms of a Hubbard model. Finally, Sec.~\ref{sec: conclusions}
contains the conclusions and the outlook.
 
\section{Multidimensional quantum capacitance}
\label{sec: MQC}

Hereafter, the scalar expression of the quantum capacitance (QC) is recalled and related, through the Hellmann-Feynman theorem, directly to the charge density and to the voltage-dependent part of the confinement potential (Subsec.~\ref{subsec:SQC}). The expressions are then generalized to the multi-dimensional case, where we introduce the concept of {\it quantum capacitance matrix} (Subsec.~\ref{subsec:VQC}). 

\subsection{Single gate voltage}
\label{subsec:SQC}

In the one-dimensional case ($M=1$), the system Hamiltonian, and therefore its eigenvalues and eigenstates,
depend on the voltage $V$ applied to a single gate. If the value of the
voltage only undergoes small deviations from the working point $V_0$, then the energy $E_k$ of the $k$-th few-particle eigenstate $\big| \psi_k \big>$ at $V = V_0 + \delta V$ can be approximated by the lowest terms of its Taylor expansion around $V_0$:
\begin{align}
E_{k} (V) \approx E_{k} (V_0) +  E_k'(V_0) \, \delta V + \frac{1}{2} E_k''(V_0)  \, (\delta V)^2 \,,
\label{Taylor energy 1 gate}
\end{align}
where $X' \equiv d X / dV$. Here and in the following, we assume that $\big| \psi_k \big>$ is non-degenerate and that $E_k'(V_0)$ and $E_k''(V_0)$ are uniquely defined. 

From Eq.~\eqref{Taylor energy 1 gate}, it follows that the dependence of the system energy on $V$ can be
recast into that of a fictitious capacitor, whose capacitance is
identified with the second derivative of the energy with respect to
voltage\cite{Gonzalez-Zalba2016a,Petersson2010a}:
\begin{align}
      C_{k} (V_0) \equiv E_k''(V_0) .
\label{quantum capacitance}
\end{align}
In the following, in order to simplify the notation, we omit to specify that the derivatives with respect to the voltage, including the QC, are always computed at $V_0$.
Combining the two equations above, the energy eigenvalue in the vicinity
of the working point can be written in the form:
\begin{align}
      E_{k} (V) \approx   \frac{1}{2} C_{k} \, (V-V_1)^2
       + \left[E_k(V_0)  -  \frac{(E'_k)^2}{2 C_k} \right] ,
      \label{capacitor energy}
\end{align}
where $V_1\equiv V_0-E'_k/C_k$ plays the role of the effective voltage applied to the second plate of the fictitious capacitor, and the term in square brackets represents an additive constant.

While the above expression relates the QC to the system energy, it might be useful to highlight its direct relation to the confinement potential, the charge density, and their derivatives with respect to $V$. In order to do so, one can start by writing the $N$-particle Hamiltonian as follows:
\begin{align}
& \hat{H} = \hat{H}_0  + \hat{W} \,, \\
& \hat{W} = \int d \boldsymbol{r} W(\boldsymbol{r}; V) \hat{n}(\boldsymbol{r}) \,,
\label{general H}
\end{align}
where $\hat{H}_0$ includes the kinetic and Coulomb-interaction terms, and the contributions to the confinement potential that do not depend on the gate potential, such as those resulting from the band offsets in a heterostructure. The quantity $W(\boldsymbol{r}; V)$ is the voltage-dependent potential generated by the metallic gate, $\hat{n}(\boldsymbol{r}) = \sum_{B} \hat{\psi}^{\dagger}_B(\boldsymbol{r}) \hat{\psi}_B(\boldsymbol{r})$ is the density operator at position $\boldsymbol{r}$, and $B$ is a band (or spin) index.

Then, applying the Hellmann-Feynman theorem \cite{Feynman39}, the first derivative of the energy eigenvalue with respect to the voltage is written as:
\begin{align}
\frac{\partial E_k}{ \partial V} = \big< \psi_k \big|  \hat{W}' \big|
\psi_k \big> =
\int\,d\boldsymbol{r}\,W'(\boldsymbol{r};V)\,n_k(\boldsymbol{r}) .
\end{align}
Here, $W'(\boldsymbol{r};V) \equiv \partial W(\boldsymbol{r};V) /
\partial V$, and $n_k(\boldsymbol{r})$ is the charge density
corresponding to the $k$-th eigenstate, given by $n_k(\boldsymbol{r}) =
  \big< \psi_k \big| \hat{n}(\boldsymbol{r})  \big| \psi_k \big>$. A further differentiation yields the QC 
associated to the $N$-particle system:
\begin{align}\label{eq:QCHF}
C_k = \int d\boldsymbol{r} \left[
W''(\boldsymbol{r};V)\,n_k(\boldsymbol{r}) +
W'(\boldsymbol{r};V)\,n'_k(\boldsymbol{r}) \right] \,.
\end{align}
If the dependence of the function $W$ on $V$ is approximately linear in the relevant range of values, the first term in the right-hand side of Eq.~\eqref{eq:QCHF} is small with respect to the second one. 

\subsection{Multiple gate voltages }
\label{subsec:VQC}

The physical assumptions underlying the following expressions of the quantum capacitance matrix are that the relevant eigenstate (typically the ground state) is nondegenerate and that the oscillations in the applied voltage are small enough to justify the truncation of the energy eigenvalue to second order in the Taylor expansion. Besides, as in the one-dimensional case, it is assumed that the dynamics induced by the oscillating voltage is adiabatic. If this is not the case, one should take into account additional contributions, related to (coherent and incoherent) nonadiabatic processes.

\subsubsection{Quantum capacitance matrix}

The above definitions of the QC can be generalized to the case where the quantum system is electrostatically coupled to a set of $M$ gates.  Their voltages $V_i$ ($i = 1, \ldots, M$) can be regarded as the components of a vector $\boldsymbol{V}$, defined in a $M$-dimensional parameter space. Along the same lines of what has been done in the one-dimensional case, one derives a QC matrix by expanding the $k$-th energy eigenvalue around the working point
$\boldsymbol{V}_0$:
\begin{align}
E_k(\boldsymbol{V}) & \approx E_k(\boldsymbol{V}_0) \!+\!
  \sum_{i=1}^M \frac{ \partial E_k  }{\partial V_i} \,\delta V_i \!+\!
\frac{1}{2} \sum_{i, j = 1}^M  \delta V_i \frac{ \partial^2 E_k 
}{\partial V_i \partial V_j}  \delta V_j \,,
\label{energy expansion multiple gates}
\end{align}
where $\boldsymbol{V} = \boldsymbol{V}_0 + \delta \boldsymbol{V}$. The Hessian matrix of $E_k$ is identified with the {\it quantum capacitance matrix} ${\bf C}_k$, whose generic element is
\begin{align}
C_{k; ij} \equiv \frac{ \partial^2 E_k  }{\partial V_i   \partial V_j} 
\,.
\label{QC multigate}
\end{align}
The QC of the system thus depends on the direction
$\boldsymbol{v} \equiv \delta \boldsymbol{V}/ |\delta
\boldsymbol{V}|=(v_1,\dots,v_M)$ of the perturbation in the voltage
space:
\begin{align}\label{eq:qcad}
C_{k; \boldsymbol{v}} = \sum_{i, j = 1}^M  C_{k; ij} v_i v_j  \,.
\end{align}

The QC matrix can also be related to the confining potential, to the charge density, and to their derivatives with respect to the gate voltages, by applying the vectorial version of the Hellmann-Feynman theorem. For an external potential $W(\boldsymbol{r};
\boldsymbol{V})$, which depends on the voltage vector $\boldsymbol{V}$,
the gradient of the energy eigenvalue is given by:
\begin{align}
\frac{\partial E_k}{\partial \boldsymbol{V}} = \big< \psi_k
\big| \frac{\partial \hat{W}}{\partial \boldsymbol{V}} \big| \psi_k
\big> = \int d \boldsymbol{r} \frac{\partial W(\boldsymbol{r};
\boldsymbol{V})}{\partial \boldsymbol{V}} n_k(\boldsymbol{r}) \,.
\end{align}
 From this it follows that the elements of the QC matrix can be written as:
\begin{align}
C_{k; ij}\!=\! \int d \boldsymbol{r} \Bigg[ \frac{\partial^2
W(\boldsymbol{r}; \boldsymbol{V})}{\partial V_i \partial V_j}
n_k(\boldsymbol{r}) \!+\! \frac{\partial W(\boldsymbol{r};
\boldsymbol{V})}{\partial V_i}   \frac{ \partial n_k(\boldsymbol{r})
}{\partial V_j}   \Bigg] \,.
\label{QC after HF multigate}
\end{align}
Although it is not obvious from the above expression, the QC matrix is symmetric (see Appendix \ref{app:symmetry proof}). Therefore, it is always possible to diagonalize it.

\subsubsection{Eigenvalues and eigenvectors}

The diagonal element $C_{k; ii}$ of the QC matrix has a
straightforward physical meaning: it corresponds to the QC associated with small changes of the potential along the coordinate axis $V_i$. However, the coordinate axes might not represent the optimal directions along which the perturbation $\delta\boldsymbol{V}$ should be applied, i.e., the directions along which the QC reaches its maximum absolute value. Instead, the
optimal direction can be identified by diagonalizing the matrix ${\bf C}_k$ and thus deriving its eigenvalues $C_{k;n}$ and normalized eigenvectors $\boldsymbol{v}_{k;n}$ ($n=1,\dots,M$). If the voltage perturbation is parallel to the $n$-th eigenvector of the matrix, $\delta \boldsymbol{V} = \boldsymbol{v}_{k;n} \,\delta V$, then the corresponding energy can be approximated as:
\begin{align}
E_k(\boldsymbol{V}) & \approx E_k(\boldsymbol{V}_0) + Q_{k;n} \, \delta
V + \frac{1}{2} C_{k;n} \, \delta V^2 \,,
\label{energy expansion multiple gates, after diagonalization}
\end{align}
where
\begin{align}
Q_{k;n} &
\equiv
\frac{\partial E_k}{\partial \boldsymbol{V}} \cdot \boldsymbol{v}_{k;n}
\,.
\end{align}
Along the same lines of the single-voltage case, for any nonzero eigenvalue $C_{k;n}$ one can write:
\begin{align}
      E_{k} (\boldsymbol{V}) \approx & \frac{1}{2} C_{k;n} \,
\left(V_{k;n} - V_{1;k; n}\right)^2
      \!+\! \left[E_k(\boldsymbol{V}_0) \!-\! \frac{(Q_{k;n})^2}{2
C_{k;n}} \right] ,
\end{align}
where $V_{k;n} \equiv \boldsymbol{V} \cdot \boldsymbol{v}_{k;n}$ and
$V_{1;k; n} =   \boldsymbol{V}_0 \cdot \boldsymbol{v}_{k;n}  - Q_{k;n}/ C_{k;n}$.
One can thus associate to each nonzero eigenvalue of the QC matrix a fictitious capacitor with capacitance $C_{k;n}$ and voltage $V_{1;k; n}$ applied to the second plate.

\begin{figure}
\centering
\includegraphics[scale=0.5]{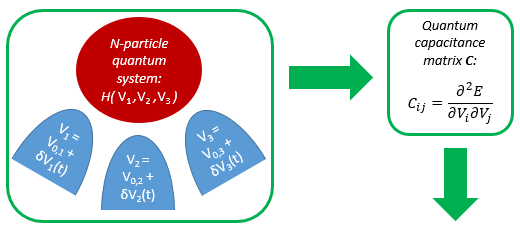}
\includegraphics[scale=0.5]{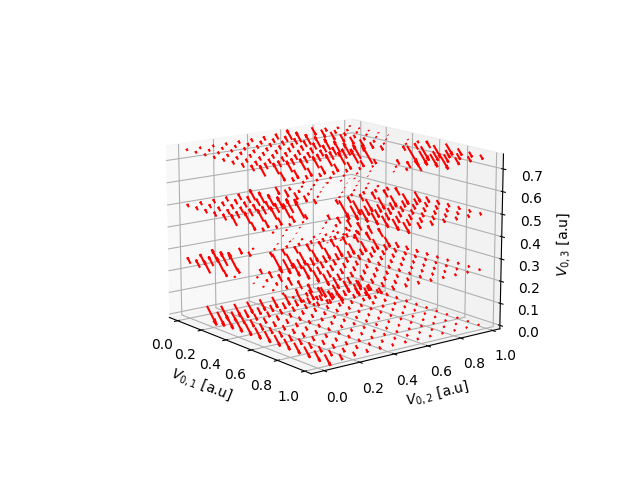}
\caption{Schematic view of the procedure based on the QC matrix. The Hamiltonian of the few-particle quantum system parametrically depends on the voltages of the $M$ metal gates (here $M=3$). The QC matrix is defined by the second partial derivatives of the relevant energy eigenvalue with respect to the voltages. From the matrix diagonalization, one can derive a vector field in the voltage space, where the direction and modulus of the field at each working point $\boldsymbol{V}_0$ respectively define the optimal combination of voltage perturbations $\delta V_i$ and the corresponding maximum of the QC.}
\label{Fig:0}
\end{figure}

Applying a generic perturbation $\delta \boldsymbol{V}  = \delta V \boldsymbol{v}$ in the voltage space causes the excitation of different normal modes of the QC matrix. The corresponding QC [Eq.~\eqref{eq:qcad}] is the weighted average of the eigenvalues $C_{k;n}$, with weights given by $p_{k;n} = (\boldsymbol{v}_{k;n} \cdot  
\boldsymbol{v})^2$, with $\sum_n p_{k; n} = 1$. It follows that the value of $C_{k; \boldsymbol{v}}$, for any $\boldsymbol{v}$, is between the lowest and the highest eigenvalues of the QC matrix at the same working point. Therefore, the optimal strategy to maximize the QC consists in applying the voltage perturbation along the direction $\boldsymbol{v}_{k;n}$ that corresponds to the eigenvalue $C_{k;n}$ having the largest modulus (Fig.~\ref{Fig:0}).

\subsubsection{Estimate of the off-diagonal matrix elements}

The QC matrix of a given system follows from the above equations, if the system Hamiltonian is known and the parametric dependence of the eigenvalues can be calculated. If this is not the case, then the matrix has to be reconstructed experimentally (fully or partially) from measurable quantities \cite{Mills2019}.  

The application of a perturbation to one of the gate voltages at a time allows to directly access the diagonal element of the QC matrix corresponding to that gate voltage. In fact, let $\boldsymbol{u}_i$ be a $M$-dimensional unit vector, with $\boldsymbol{u}_1 = (1,0,0, \ldots, 0)$, $\boldsymbol{u}_2 = (0,1,0, \ldots, 0)$, and so on. From Eq.~\eqref{eq:qcad}, one can see that, if $\boldsymbol{v} = \boldsymbol{u}_i$, then the corresponding QC is $C_{k; \boldsymbol{v}} = C_{k; ii}$. Therefore, with single-gate perturbations one can access the diagonal elements of the QC matrix.

Any off-diagonal element $C_{k; ij}$ can be estimated by following a simple two-step procedure. In the first step, one measures the diagonal terms $C_{k; ii}$ and $C_{k; jj}$, by following the procedure described above. In the second step, the perturbation is applied along the diagonal direction: $\boldsymbol{v} = \frac{1}{\sqrt{2}} (\boldsymbol{u}_i+ \boldsymbol{u}_j)$. The QC along this direction is given by [Eq.~\eqref{eq:qcad}]:
\begin{align}
C_{k;\boldsymbol{v}}  = \frac{1}{2}  \left( C_{k;ii} +    C_{k;jj} \right)  +    C_{k;ij}  \,.
\end{align}
The off-diagonal element $C_{k; ij} = C_{k; ji}$ can thus be derived as the
difference between the above QC and the average of the diagonal elements that have been measured in the first step. The knowledge of the off-diagonal elements allows one to identify the optimal voltage direction within the two-dimensional subspace spanned by $\boldsymbol{u}_i$ and $\boldsymbol{u}_j$. This can be done by diagonalizing the sub-matrix of the QC matrix defined by the $i$-th and $j$-th lines and columns.

Analogously, one can identify the optimal voltage direction within any subensemble formed by $n$ of the $M$ gate voltages or, equivalently, within the subspace spanned by $n$ vectors $\boldsymbol{u}_i$ in the voltage space. Therefore, either by computing the parametric dependence of $E_k$ on $\boldsymbol{V}$, or by repeating the above procedure for each couple of gates $(i,j)$ (with $i < j$), one can reconstruct the whole QC matrix and the derive its eigenvalues and eigenvectors by numerical diagonalization.

\section{Application to a quantum-dot array}
\label{sec: Hubbard}

In order to illustrate the possible applications of the theory discussed so far, we consider an array of $M>1$ tunnel-coupled quantum dots, modeled by means of a Hubbard Hamiltonian. We denote the number of electrons populating the system as $N$. Because of the presence of tunneling between the dots, the gate voltages allow to modulate and probe interacting $N$-particle states which are, in general, delocalized over the whole array. We first derive several general properties of the QC matrix for such model (Subsec.~\ref{subsec:HME}), and then show the application of the theory to a few representative cases corresponding to $M=3$ (Subsec.~\ref{subsec:HMR}). With respect to the results presented so far, the ones reported in the present Section are based on further assumptions on the system properties. In particular, the expressions derived in Subsec.~\ref{subsec:HME} present the same range of validity of the underlying Hubbard model, essentially based on the assumptions that: the on-site repulsion is much larger than the tunneling, the excited orbitals of the three dots and the inter-site repulsion do not to play any role. As to the numerical results discussed in Subsec.~\ref{subsec:HMR}, they are based on the further, simplifying assumption that the gate voltages affect linearly the on-site energies and leave unaffected both the tunneling amplitude and the on-site repulsion.

\subsection{Quantum capacitance in the Hubbard model}
\label{subsec:HME}

The quantum-dot array is represented in terms of a Hubbard model with one orbital per dot (site), whose Hamiltonian reads:
\begin{align}
\hat{H}   \equiv \sum_{i=1}^M  \left[ \epsilon_i \hat{n}_i   + \frac{ U }{ 2 } \hat{n}_i \left( \hat{n}_i - 1 \right)  \right]  +    \hat{H}_T  \,,
\label{Hubbard model}
\end{align}
where $\epsilon_i$ is the onsite energy of the $i$-th quantum dot, $U$ is the onsite Coulomb-interaction energy, and $\hat{n}_i = \sum_{\sigma}  \hat{c}^{\dagger}_{i, \sigma} \hat{c}_{i, \sigma} $ is the number operator of the $i$-th quantum dot ($\sigma = \uparrow, \downarrow$ labels the two possible single-electron spin orientations); $\hat{H}_T$ is the hopping Hamiltonian, which has the general form
\begin{align}
\hat{H}_T =  \sum_{i = 1}^M \sum_{j = 1}^M T_{i,j} \sum_{\sigma} \hat{c}^{\dagger}_{i, \sigma} \hat{c}_{j, \sigma}    \,. 
\end{align}

In general, all parameters $\epsilon_i$, $U$ and $T_{i,j}$ depend on the gate voltages, as they are matrix elements of either voltage-dependent ($\epsilon_i$, $T_{i,j}$) or voltage-independent ($U$) operators between basis functions which do depend on the gate voltages in all cases. Application of the Hellmann-Feynman theorem to the $N$-electron eigenstate $\big| \psi_k \big>$ with energy $E_k$ yields the equation:
\begin{align}
\frac{\partial E_k}{\partial V_i}  & = \sum_{l = 1}^M   \frac{\partial \epsilon_l}{\partial V_i}   \big< \hat{n}_l \big>_k + \frac{\partial U}{\partial V_i} \left( \sum_{l = 1}^M  \big<  \hat{n}^2_l   \big>_k  - N \right)    \nonumber \\
& \quad  + \sum_{l = 1}^M \sum_{l' = 1}^M \frac{ \partial T_{l,l'} }{\partial V_i } \sum_{\sigma} \big< \hat{c}^{\dagger}_{l, \sigma} \hat{c}_{l', \sigma}  \big>_k      \,,
\label{Hellmann-Feynman Hubbard general case}
\end{align}
where $\big< \ldots \big>_k \equiv \big< \psi_k \big| \ldots \big| \psi_k \big>$.
By further differentiating, one obtains the following expression for the elements of the QC matrix:
\begin{align}
C_{k; ij} &   =    \sum_{l = 1}^M   \frac{\partial^2 \epsilon_l}{\partial V_i \partial V_j}   \big< \hat{n}_l \big>_k  + \sum_{l = 1}^M   \frac{\partial \epsilon_l}{\partial V_i}  \frac{ \partial  \big< \hat{n}_l \big>_k }{\partial V_j} \nonumber \\
& \quad  + \frac{\partial^2 U}{\partial V_i \partial V_j} \left( \sum_{l = 1}^M  \big<  \hat{n}^2_l   \big>_k  - N \right)   
+  \frac{\partial U}{\partial V_i}   \sum_{l = 1}^M  \frac{ \partial  \big<  \hat{n}^2_l   \big>_k }{\partial V_j}     \nonumber \\
& \quad  + \sum_{l = 1}^M \sum_{ l' = 1}^M   \sum_{\sigma}  \Bigg( \frac{ \partial^2 T_{l,l'} }{\partial V_i \partial V_j } \big< \hat{c}^{\dagger}_{l, \sigma} \hat{c}_{l', \sigma}  \big>_k   \nonumber \\
& \quad   + \frac{ \partial T_{l,l'} }{\partial V_i } \frac{ \partial \big< \hat{c}^{\dagger}_{l, \sigma} \hat{c}_{l', \sigma}  \big>_k }{\partial V_j }  \Bigg)   \,.
\label{eq:qc1 general case}
\end{align} 

For illustrative purposes, we now make some simplifying assumptions. First, we assume that the onsite energy of each dot is linearly dependent on the voltage applied to the corresponding gate, and independent of the others:
\begin{equation}\label{eq:linearity}
     \frac{\partial \epsilon_i}{\partial V_j} = \alpha_i \,\delta_{i,j}\,.
\end{equation}
Second, variations of the gate voltages are assumed not to significantly affect $T_{i, j}$ and $U$. These assumptions are introduced in order to simplify the discussion, but are not required within the proposed approach and can be removed in order to include further aspects in the analysis of the quantum-dot array (see Ref.~\onlinecite{e25010082} for the study of a case where the tunneling is affected by gate voltages).

Under the above assumptions, Eq.~\eqref{Hellmann-Feynman Hubbard general case} simplifies as
\begin{align}
\frac{\partial E_k}{\partial V_i} = \alpha_i \big< \psi_k \big|
\frac{\partial \hat{H}}{\partial \epsilon_i} \big| \psi_k
\big> = \alpha_i \big< \hat{n}_i \big>_k \,,
\label{Hellmann-Feynman Hubbard}
\end{align}
and the elements of the QC matrix become
\begin{align}\label{eq:qc1}
C_{k; ij} = \alpha_i \frac{\partial \big< \hat{n}_i \big>_k}{\partial V_j} =
\alpha_j \frac{\partial \big< \hat{n}_j \big>_k}{\partial V_i} \,.
\end{align}
The matrix element $C_{k; ij}$ is thus related to the dependence of the average occupation in dot $i$ ($j$) on the voltage $V_j$ ($V_i$). Even if the Hamiltonian terms involving the operator $\hat{n}_i$ do not depend on $V_j$ for $j \neq i$ (as per the above assumptions), the expectation value $\big< \hat{n}_i \big>_k$ does, because of the quantum correlations induced by the Hamiltonian in its eigenstates $\big| \psi_k \big>$. Therefore, in general, the off-diagonal elements of the QC matrix differ from zero.

From the Hellmann-Feynman theorem, one can also derive the following sum
rule:
\begin{align}
\sum_{i = 1}^M \frac{1}{\alpha_i} \frac{\partial E_k}{\partial V_i} = N
\,,
\label{gradient sum rule}
\end{align}
where $N$ is the number of particles.
 From this it follows that the QC matrix always has a zero eigenvalue. In fact, by combining Eqs.~\eqref{Hellmann-Feynman Hubbard} and \eqref{gradient sum rule}, one obtains:
\begin{align}
     \sum_{j=1}^M \frac{1}{\alpha_j} C_{k;ij} = \frac{\partial}{\partial
V_i} N = 0,
\end{align}
for any $i=1,\dots,M$. As a consequence, for any $M$ and $N$, the normalized vector
$\boldsymbol{v} = A (\alpha_1^{-1},\dots,\alpha_M^{-1}) $ (where
$A^{-2}=\sum_{i=1}^M\alpha_i^{-2}$) is an eigenvector of the QC matrix, with zero eigenvalue. This physically corresponds to the case where all the energies $\epsilon_i$ undergo the same perturbation, producing a rigid shift of all the energy eigenvalues that -- in view of Eq.~\eqref{eq:linearity} -- is linear in the voltage. Such a zero-capacitance mode does not contribute to the QC $C_{k;\boldsymbol{v}}$ for any $\boldsymbol{v}$, and can be disregarded in the following analysis.

\subsection{QC measurements in the case of a few-dot array}
\label{subsec:HMR}

In the present Subsection, we apply the results derived in the previous one to the investigation of a few representative cases, defined by specific numbers of sites ($M$) and of particles ($N$). Each eigenstate is characterized by defined value of the total spin $S$, because the total spin operator ${\bf S}^2$ commutes with the Hubbard Hamiltonian. 

The dependence of the QC on the working point and on the direction of the voltage oscillations defines a pattern, which presents $N$- and $S$-dependent features and can thus be exploited in order to infer the values of these quantum numbers. 

A preliminary characterization of these patterns is derived hereafter on the basis of a semiclassical picture, and then integrated by a fully quantum-mechanical approach. The semiclassical picture identifies the regions in the voltage space where the QC is expected to take significant values. The quantum-mechanical approach determines the actual values in such regions of the QC, and the directions along which the voltage oscillations should be induced in order to maximize its absolute value.

\subsubsection{Semiclassical picture}

\paragraph{General aspects.}
In general, the QC is expected to take significant values at the boundaries between charge stability regions (CSRs) in the voltage space, where small fluctuations in $\boldsymbol{V}$ can induce significant rearrangements of the charge [Eq.~\eqref{eq:qc1}]. These boundaries can be identified within a semiclassical picture, obtained by removing the hopping term from the Hamiltonian [Eq.~\eqref{Hubbard model}], and thus the quantum fluctuations in the dot occupations from the eigenstates. The energies for each set of voltages $\boldsymbol{V}$ and dot occupations
$\boldsymbol{n} = (n_1,\dots,n_M)$ are given by:
\begin{align}
E_{\boldsymbol{n}}(\boldsymbol{V})   = \sum_{i=1}^M \left[ \alpha_i V_i 
n_i + U \frac{n_i (n_i-1)}{2} \right] \,.
\label{classical energy}
\end{align}
These energies approach the eigenvalues of the Hubbard Hamiltonian in the limit where the biases induced by the occupation-conserving terms, i.e. $|\epsilon_i-\epsilon_j|$ or
$|\epsilon_i-\epsilon_j \pm U|$, are much larger then the amplitude of the (direct or indirect) particle hopping between sites $i$ and $j$. Each CSR is identified by the occupations $\boldsymbol{n}$ that, for each $\boldsymbol{V}$ belonging to the CSR, minimize the energy Eq.~\eqref{classical energy}.

One can show that the degeneracy between the energies of any two configurations $\boldsymbol{n}$ and $\boldsymbol{n}'$ occurs at hyperplanes in $\mathbb{R}^M$, specified by the equation
\begin{align}
\pi_{A, \boldsymbol{B}} = \left\{ \boldsymbol{V} \in \mathbb{R}^M :  A +
\boldsymbol{B} \cdot \boldsymbol{V} = 0  \right\}  \ \ \
(|\boldsymbol{B}|=1)\,,
\label{hyperplane}
\end{align}
where $A$ is an offset and $\boldsymbol{B}$ is a $M$-dimensional vector perpendicular to the hyperplane. All these hyperplanes are parallel to the direction $\boldsymbol{D}=(\alpha_1^{-1},\dots,\alpha_M^{-1})$, corresponding to the zero eigenvalue of the QC matrix (see Appendix \ref{app: hyperplanes}). Therefore, any two adjacent CSRs are separated by a part of a hyperplane, delimited by intersections with other hyperplanes.

Whenever one moves from a CSR to an adjacent one by crossing a hyperplane $\pi$, e.g. by varying the working point in the voltage space starting at $\boldsymbol{P}_0$ and moving along the line
\begin{align}
l_{\boldsymbol{C}, \boldsymbol{P}_0} = \left\{ t \in \mathbb{R} : \boldsymbol{V}_0 =
\boldsymbol{P}_0 + t \boldsymbol{C} \right\} \ \ \ (|\boldsymbol{C}|=1)\,,
\label{hyperline}
\end{align}
one expects a peak in the QC at the intersection between $l_{\boldsymbol{C}, \boldsymbol{P}_0}$ and $\pi$. Intuitively, the highest peak is expected to be achieved if the voltage oscillation $\delta \boldsymbol{V}$ is perpendicular to the hyperplane ($\boldsymbol{v} \parallel \boldsymbol{B}$). In fact, any component of the oscillation parallel to the plane is expected to induce no rearrangement of the charge, and thus to provide no contribution to the QC.

\paragraph{Three-dot array.}
Many of the relevant aspects of the approach can be illustrated within a simple three-dot array ($M=3$). In this case, the boundaries between stability regions are given by planes perpendicular to one of the following three vectors:
\begin{align}
& \boldsymbol{B}_1   = \frac{1}{\sqrt{\alpha_1^{-2}+\alpha_2^{-2}}}
\left(\frac{1}{\alpha_1},-\frac{1}{\alpha_2},0\right) \,, \nonumber\\
& \boldsymbol{B}_2   = \frac{1}{\sqrt{\alpha_2^{-2}+\alpha_3^{-2}}}
\left(0,\frac{1}{\alpha_2},-\frac{1}{\alpha_3}\right) \,, \nonumber\\
& \boldsymbol{B}_3   = \frac{1}{\sqrt{\alpha_1^{-2}+\alpha_3^{-2}}}
\left(-\frac{1}{\alpha_1},0,\frac{1}{\alpha_3}\right) \,.
\end{align}
Physically, these directions correspond to transitions that involve a charge transfer respectively between the first and the second dot, the second and the third, the first and the third. The additive term $A$ is either 0 or $\pm U$, depending on whether or not the transition between CSRs implies a change in the number of doubly occupied dots.

\begin{figure}
\centering
\includegraphics[scale=0.5]{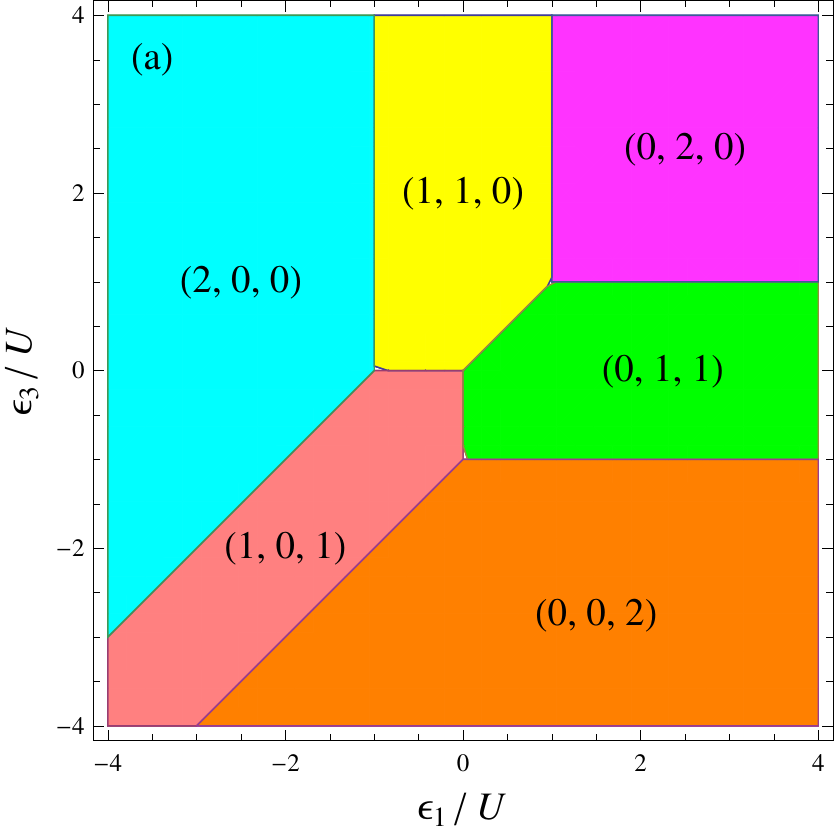}
\includegraphics[scale=0.5]{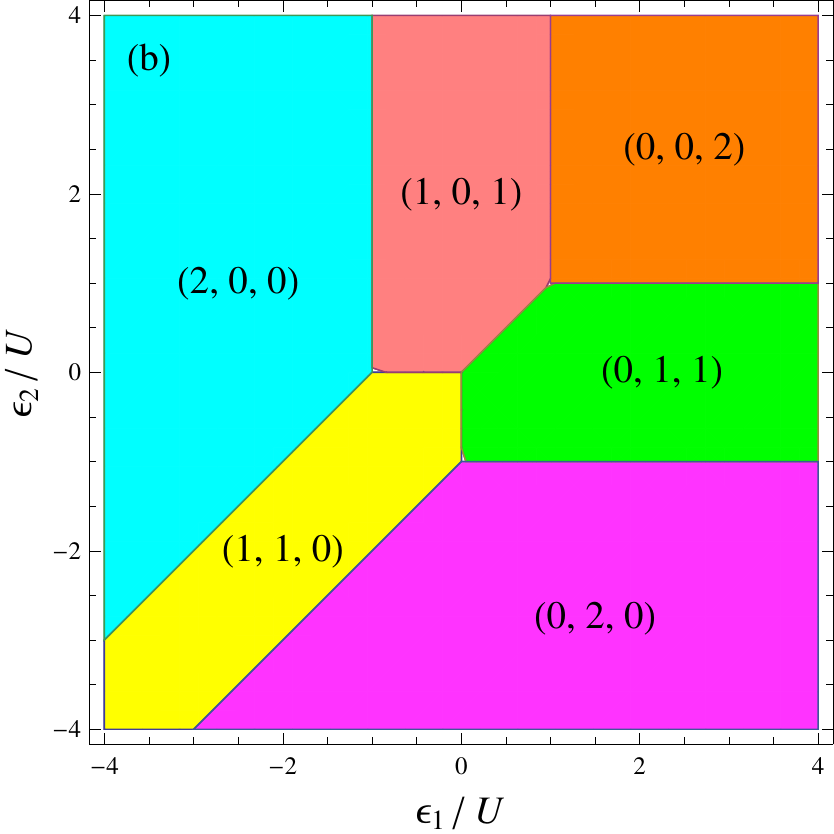}
\includegraphics[scale=0.5]{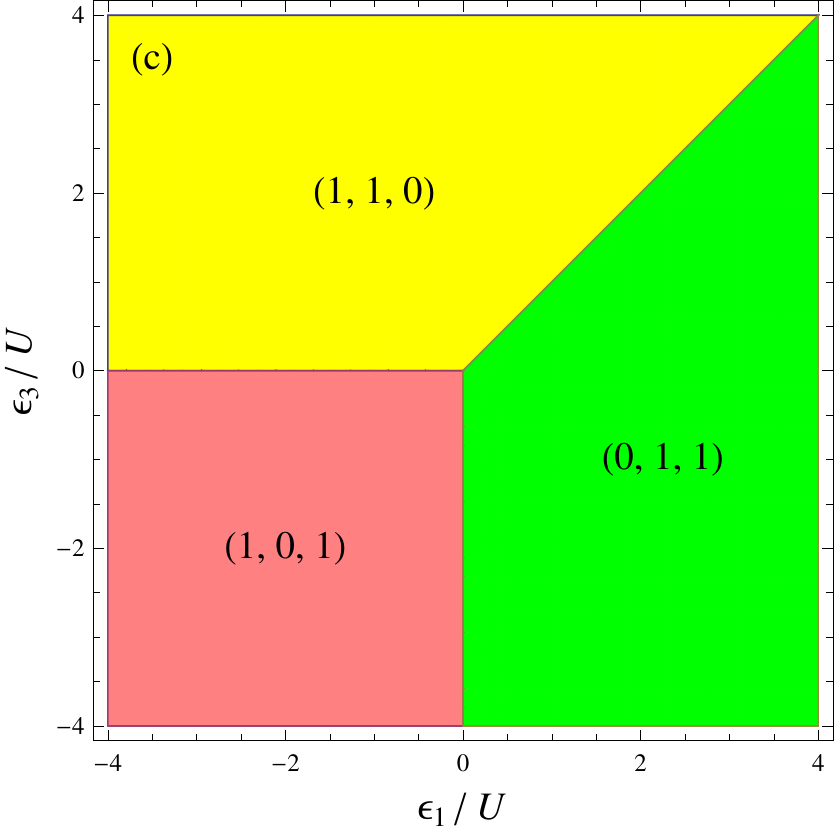}
\includegraphics[scale=0.5]{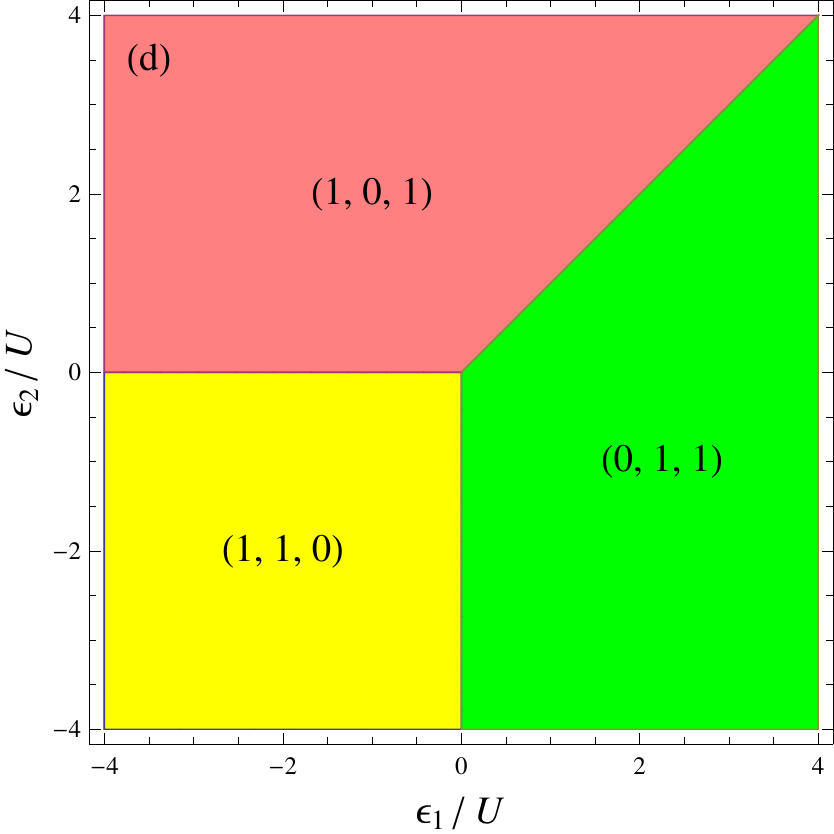}
\caption{Charge stability regions in the coordinate planes $\epsilon_2 = 0$ (a, c) and $\epsilon_3 = 0$ (b, d), for the spin-singlet (a, b) and spin-triplet (c, d) ground states of $N=2$ particles in a linear array formed by $M=3$ quantum dots. Each region is denoted by the corresponding occupations of the three dots. }
\label{Fig:1}
\end{figure}
\begin{figure}
\centering
\includegraphics[scale=0.5]{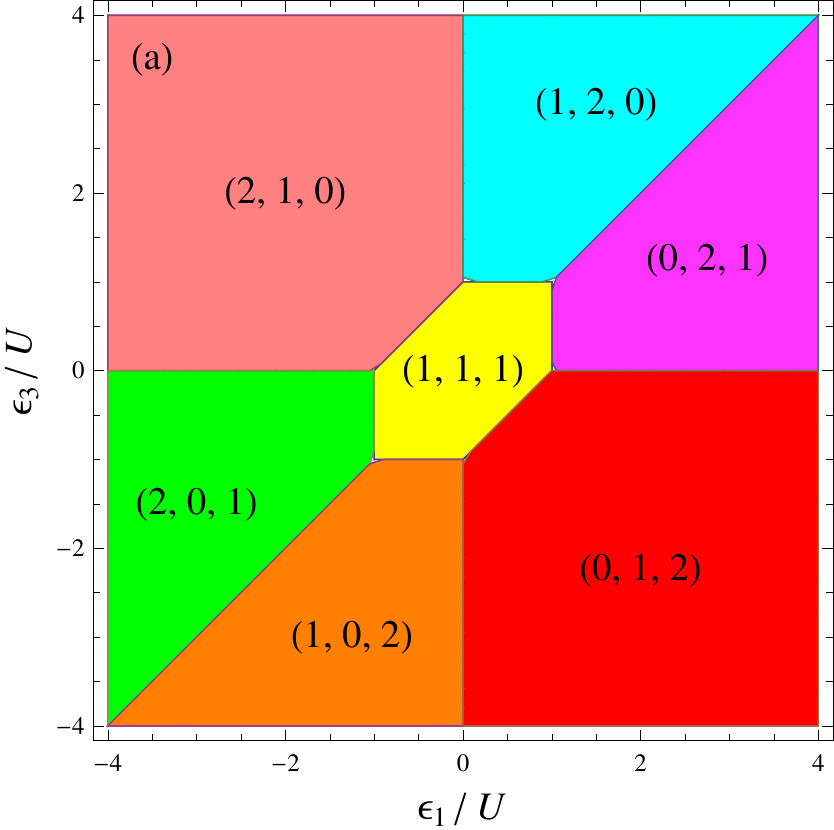}
\includegraphics[scale=0.5]{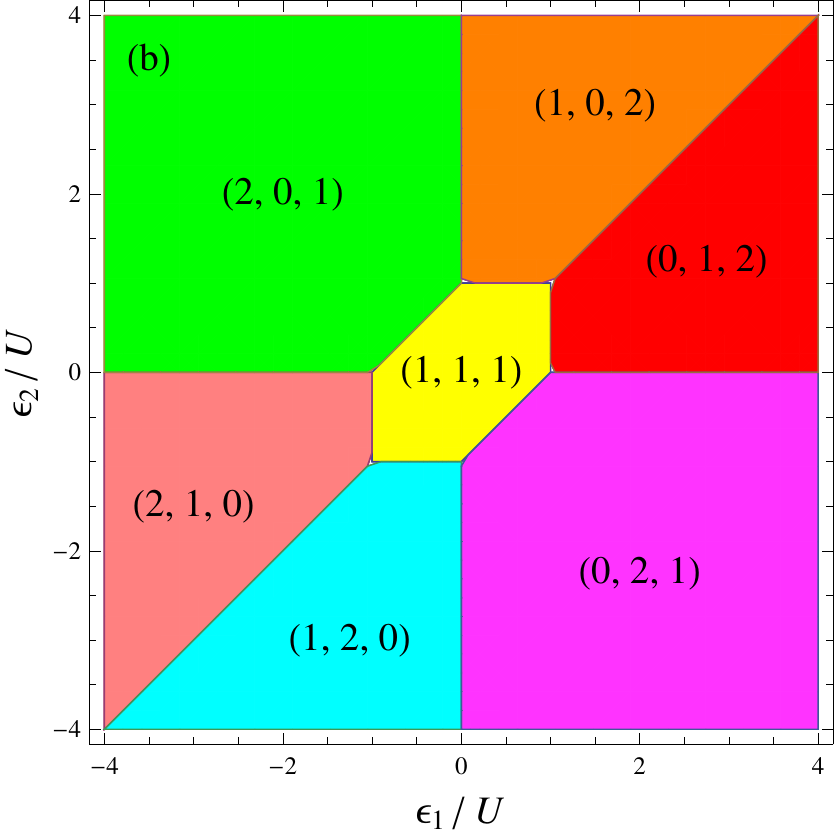}
\caption{Charge stability regions in the coordinate planes $\epsilon_2 = 0$ (a) and $\epsilon_3 = 0$ (b), for the spin-doublet ground state of $N=3$ particles in a linear array formed by $M=3$ quantum dots. Each region is denoted by the corresponding occupations of the three dots.}
\label{Fig:2}
\end{figure}

If the array is occupied by two or three particles, the ground state
corresponds to a spin singlet ($S=0$) or doublet ($S=1/2$), respectively. The intersections between the charge stability regions, identified by the dot occupations $\boldsymbol{n} = (n_1,n_2,n_3)$, and the coordinate planes $\epsilon_2=0$ and $\epsilon_3=0$ are reported in Figs.~\ref{Fig:1}(a, b) and \ref{Fig:2} respectively for the cases of $N=2$ and $N=3$. In both cases, the pattern corresponding to the plane $\epsilon_1=0$ is equivalent, up to a permutation of $n_1$ and $n_3$, to that displayed for $\epsilon_3=0$. The pattern corresponding to any plane $\epsilon_i = K$ is obtained from that for $\epsilon_i = 0$, through a translation by a quantity $K$ along the other two coordinate axes.

The QC can be used to discriminate between quantum states, if it takes significantly different values for such states in some region in the voltage space. The present semiclassical picture allows us to anticipate an important conclusion, namely that the boundaries display non-overlapping regions for states that differ either in the particle number $N$ or in the value of the total spin $S$. 
In order to assess the particle-number discrimination, one can compare the corresponding panels in Figs.~\ref{Fig:1}(a, b) and \ref{Fig:2}. In fact, the boundaries that characterize the two patterns only overlap in three points: apart from these, any working point $\boldsymbol{V}_0$ that belongs to one boundary in one of the two patterns is thus in principle suitable for discriminating between the two values of $N$. 

In Fig.~\ref{Fig:1}(c) and (d) we show the CSRs for $M=3$ and $N=2$ with the same conventions as in panels (a) and (b), respectively, but in the situation where double occupancy of a site is not allowed. This occurs due to Pauli's exclusion principle if the state of the system is a triplet ($S=1$), rather than a singlet, i.e., it is in a spin-polarized state. The comparison between the CSRs for $S=0$ and $S=1$ shows that total-spin discrimination is feasible. The CSR boundaries that are either added or removed in passing from one spin state to the other precisely define the working points $\boldsymbol{V}_0$ that are suitable for the spin discrimination. In the case of $M=3$ and $N=3$, the CSR diagram for the quadruplet ($S = 3/2$) states is a trivial one, since only the configuration $(1,1,1)$ is allowed in the whole parameter space, in marked contrast with the CSRs for the doublet which are displayed in Fig.~\ref{Fig:2}. Again, the difference between the CSRs in the two cases enables total-spin discrimination.

We note that this represents an extension to larger dot and particle numbers, and to different spin states, of the approaches that are currently used in order to readout the spin qubits through a spin-to-charge conversion or the Pauli spin blockade.

\subsubsection{Quantum-mechanical approach}

While the semiclassical picture allows the identification of the
boundaries between CSRs and suggests the optimal directions along which the voltage oscillation should take place in the $M$-dimensional voltage space, quantitative estimates require the quantum mechanical approach, i.e., the diagonalization of the Hamiltonian and the derivation of the QC matrix from the eigenenergies. In particular, this provides: a validation of the qualitative picture derived from the the semiclassical approach; a qualitative integration of such picture, for example at the intersection between three planes; a quantitative estimate of the QC at the relevant working points.

Differently from the general semiclassical consideration reported above, the details of the quantum-mechanical results depend on the hopping Hamiltonian $\hat{H}_T$, which encodes the geometry and the coordination of the quantum-dot array. For illustrative purposes, in the following we consider a one-dimensional arrangement of $M=3$ quantum dots, with only nearest-neighbour tunneling, i.e.,
\begin{align}
\hat{H}_T & =  T \sum_{i=1}^{2} \sum_{\sigma}\left( \hat{c}^{\dagger}_{i,
\sigma} \hat{c}_{i+1, \sigma} +  \hat{c}^{\dagger}_{i + 1, \sigma}
\hat{c}_{i , \sigma}\right) \nonumber \\
& \quad + \chi  T \left( \hat{c}^{\dagger}_{3,
\sigma} \hat{c}_{1, \sigma} +  \hat{c}^{\dagger}_{1, \sigma}
\hat{c}_{ 3, \sigma}\right) 
\,.
\label{H_T}
\end{align}
Here, setting $\chi = 0 $ yields an open chain, while setting $\chi = 1 $ yields a closed ring, since it enables tunneling between the first and the last site. In the following, we will consider both cases in order to show typical situations where different geometries (encoded in different tunneling Hamiltonians) give qualitatively different features in the QC.

\paragraph{Validation of the semiclassical picture.}
The semiclassical picture suggests that a high value of the QC might be obtained by choosing a working point that lies at the boundary hyperplane between two CSRs and by varying the potentials along the direction perpendicular to such hyperplane. The validity of such intuition can be verified by comparing the QC $C_{1;\boldsymbol{v}}$ [Eq.~\eqref{eq:qcad}] for $\boldsymbol{v} \parallel \boldsymbol{B}$ with the lowest and largest-modulus eigenvalue $C_{1;1}$ of the QC matrix, as a function of the working point along the line perpendicular to the hyperplane [$\boldsymbol{C}=\boldsymbol{B}$ in Eq.~\eqref{hyperline}]. 

\begin{figure}
\centering
\includegraphics[scale=0.5]{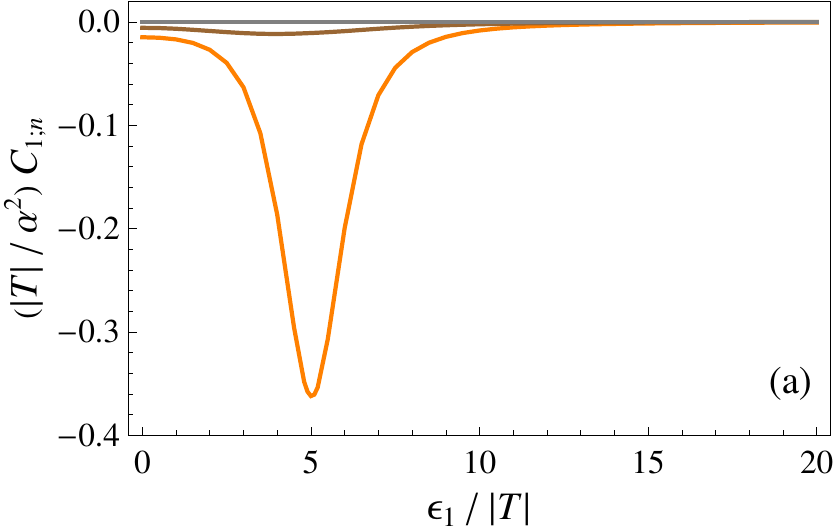}
\includegraphics[scale=0.5]{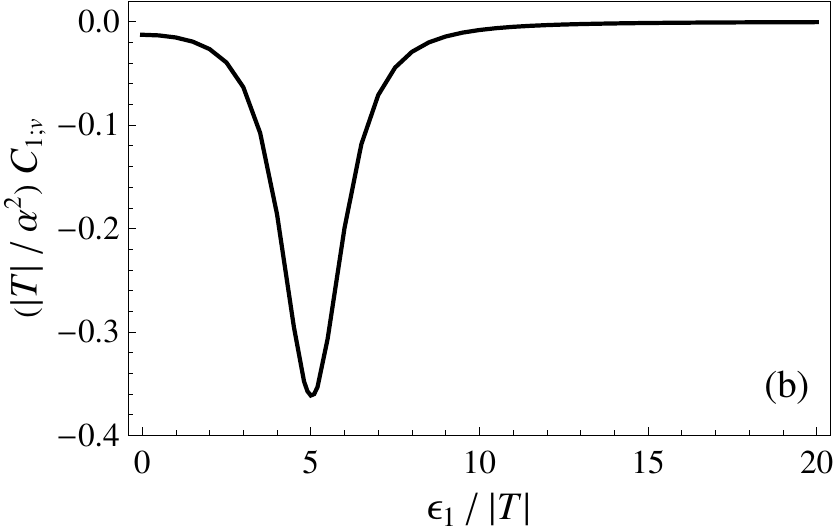}
\includegraphics[scale=0.5]{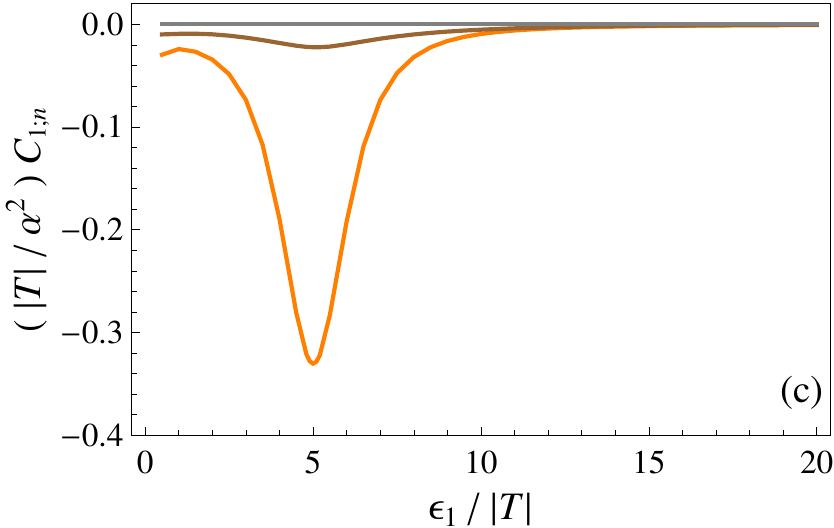}
\includegraphics[scale=0.5]{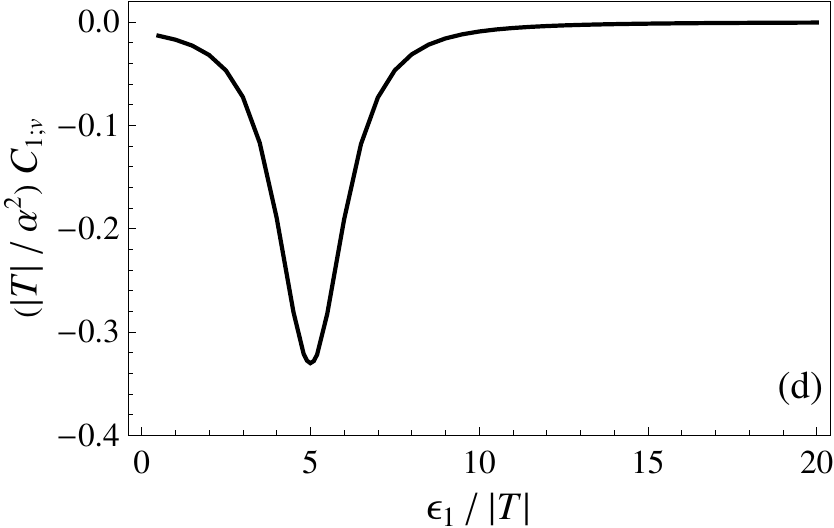}
\caption{Comparison between the eigenvalues of the QC matrix (a, c) and the scalar QC along the direction identified through the semiclassical picture (b, d), for the Hubbard model with $M=3$ quantum dots and $N=3$ particles, in the cases of an open chain (a, b) and of a closed ring (c, d). Both the direction along which the working point varies ($\boldsymbol{C}$) and the direction along which the voltage oscillates ($\boldsymbol{v}$) are parallel to $\boldsymbol{B}_1$; specifically, $\boldsymbol{\epsilon} = \epsilon_1 (1, -1, 0)$ and $\boldsymbol{v} = 2^{-1/2} (1,-1,0)$. The boundary between the $(1,1,1)$ and $(0,2,1)$ stability regions is crossed at $\epsilon_1 = U/2$. The calculations related to these graphs were done for $T < 0$, $U = 10 |T|$, and for $\alpha_i \equiv \alpha \, \forall i$. }
\label{Fig:3}
\end{figure}

Such comparison is reported in Fig.~\ref{Fig:3} in a representative case, namely for $N=3$ and for the boundary between the CSRs $\boldsymbol{n} =(1,1,1)$ and $\boldsymbol{n} =(0,2,1)$ [see Fig.~\ref{Fig:2}(b)]. We report our results both for the open chain [panels (a) and (b)] and for the closed ring [panels (c) and (d)]. In both cases, the quantum calculations fully confirm the expectation based on the semiclassical picture: the QC obtained for a direction $\boldsymbol{v}$ of the voltage oscillation orthogonal to the separation plane between the two CSRs coincides, to a good approximation, with the largest-modulus eigenvalue of the QC matrix. This also results from the eigenvector $\boldsymbol{v}_{1;1}$ of the QC matrix, which is actually oriented along the direction $\boldsymbol{B}_1$ at the boundary hyperplane. The other two eigenvalues, corresponding to orientations along the boundary plane, are exactly or approximately zero [see Fig.~\ref{Fig:3}(a) and \ref{Fig:3}(c)]. Other investigated cases (not reported) provide converging indications.

In this particular case, the difference in the two considered geometries has a very small impact on the QC, as can be seen by comparing panels (a) and (b), respectively, with panels (c) and (d). This can be understood by noticing that the dominant process determining the considered transition between charge configurations $(1,1,1)$ and $(0,2,1)$ is the tunneling between sites $1$ and $2$, which is the same in the two geometries that we are comparing. The small differences between the two cases are due to higher-order tunneling processes, which are particularly weak close to the boundary between the CSRs (i.e., $\epsilon_1 = U/2$). We will later show a case where the transition is such that geometry has a relevant impact on the QC. For the case at hand, in Appendix \ref{app:analytical1} we present an analytically solvable reduced model, which reproduces the main features of the QC displayed in Fig.~\ref{Fig:3}.

\begin{figure}
\centering
\includegraphics[scale=0.5]{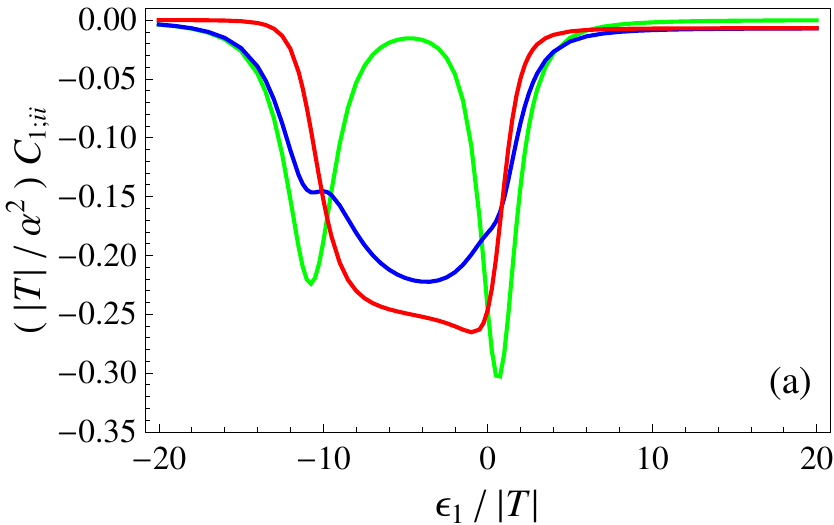}
\includegraphics[scale=0.5]{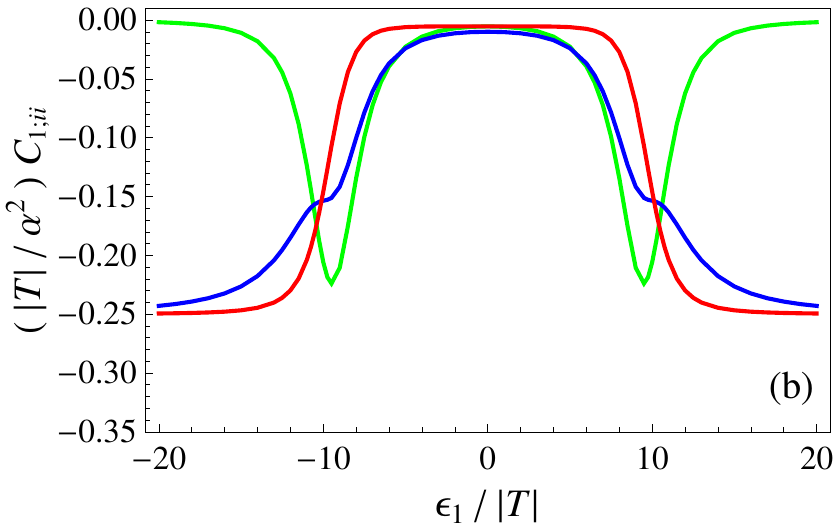}
\includegraphics[scale=0.5]{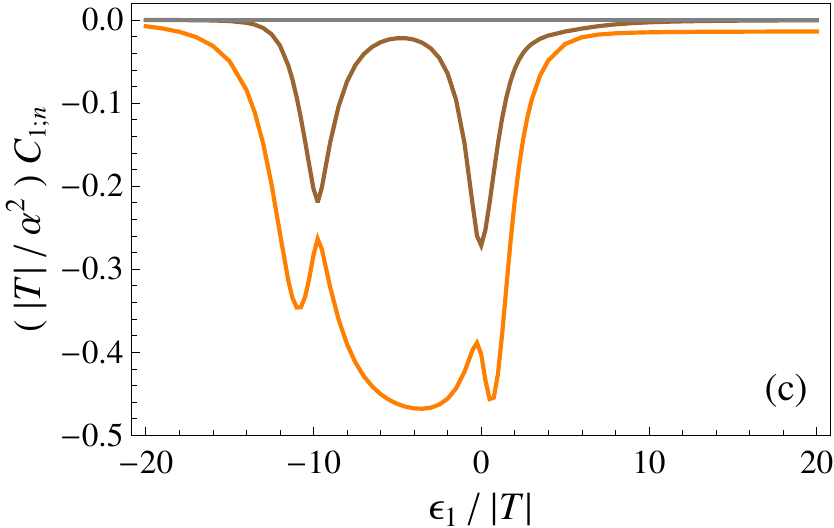}
\includegraphics[scale=0.5]{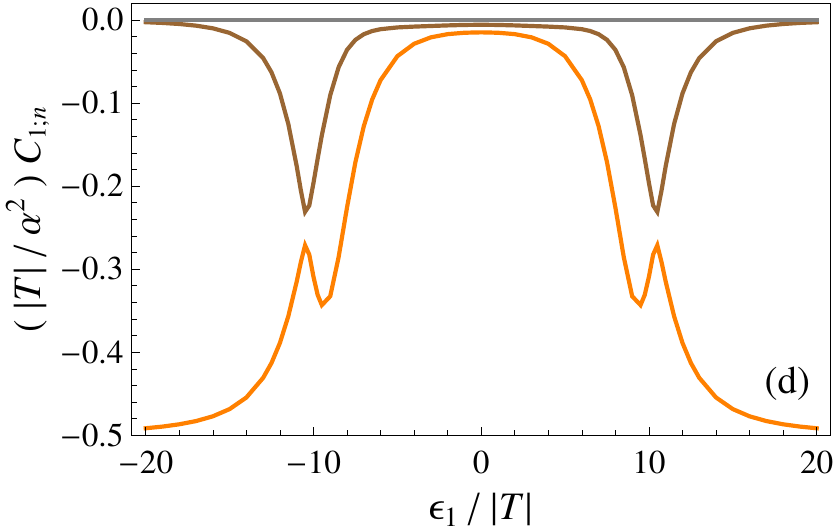}
\includegraphics[scale=0.5]{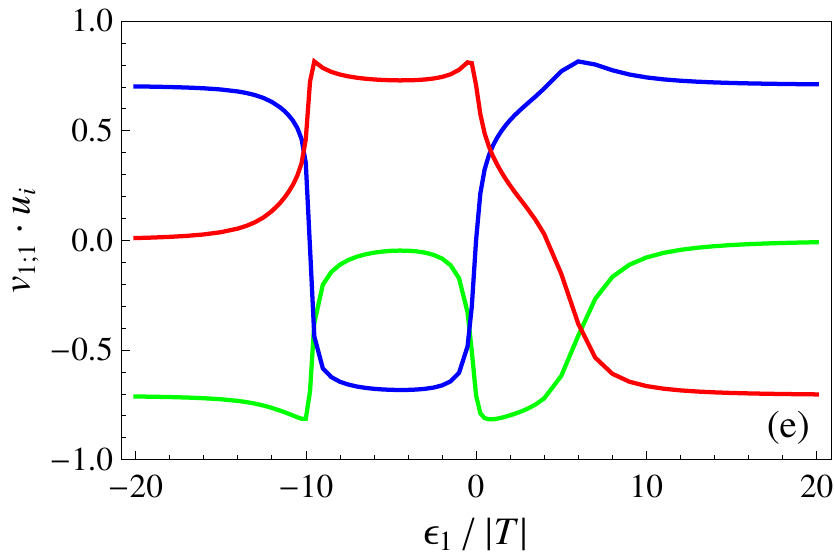}
\includegraphics[scale=0.5]{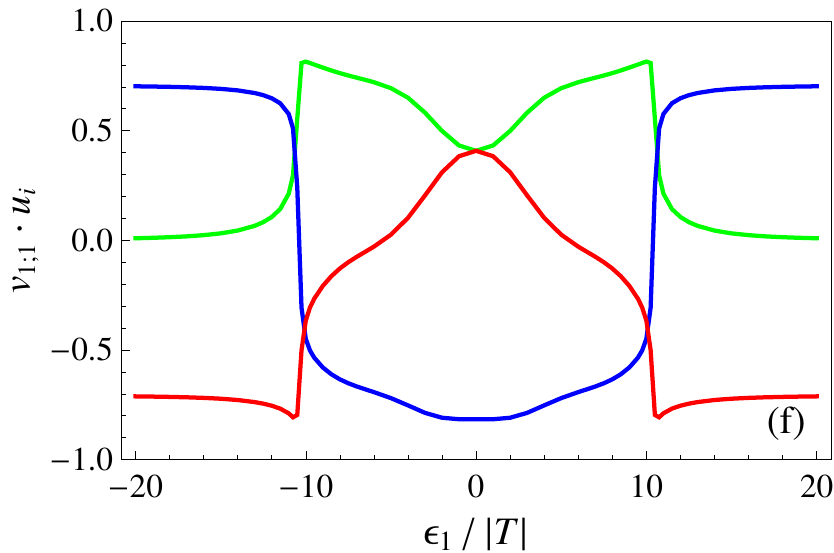}
\caption{Several quantities related to the QC matrix of the Hubbard-model ground state for $M=3$ and $N=2$ [panels (a), (c), (e)] or $N=3$ [panels (b), (d), (f)], in the case of an open chain. The calculations are done for $U = 10 |T|$, and $\alpha_i \equiv \alpha \,\, \forall i$. The various quantities are plotted as functions of $\epsilon_1$, the working point being $\boldsymbol{\epsilon} = (\epsilon_1, 0, 0)$. Panels (a) and (b): diagonal elements $C_{1; ii}$; panels (c) and (d): eigenvalues $C_{1; n}$; panels (e) and (f): components $\boldsymbol{v}_{1;1} \cdot \boldsymbol{u}_i$ of the QC eigenvector $\boldsymbol{v}_{1;1}$ corresponding to the lowest eigenvalue [orange curves in panels (c) and (d)] along the coordinate unit vectors $\boldsymbol{u}_i$. QCs are displayed in dimensionless form (see the label of vertical axes). The colors of the curves in panels (a), (b), (e), and (f) identify the corresponding value of $i$ according to the following color code: green $\rightarrow i=1$, blue $ \rightarrow i=2$, red $\rightarrow i=3$. 
}
\label{Fig:4}
\end{figure}

\paragraph{Qualitative integration of the semiclassical picture.}
The quantum approach allows a full characterization of the multidimensional QC, especially at the boundaries between CSRs, which are the most significant regions in the $\boldsymbol{V}$ space. It provides the optimal directions for the voltage oscillations and the corresponding largest-modulus eigenvalues of the QC matrix. This is especially important close to the intersection between different boundary planes, where the classical expectation might be wrong or even undefined. These aspects are discussed in some detail hereafter for the representative cases of $M=3$ and $N=2,3$ in an open chain (Fig.~\ref{Fig:4}). All quantities are plotted as a function of the position of the working point $\boldsymbol{V}_0$ along the $V_1$ axis.

In the case of $N=2$, the working-point axis lies in the CSRs $(2,0,0)$ and $(0,1,1)$, respectively for $\epsilon_1 = \alpha_1 V_1 < -U$ and $V_1>0$ (Fig.~\ref{Fig:1}). In these regions, both the diagonal elements of the QC matrix [Fig.~\ref{Fig:4}(a)] and its eigenvalues [Fig.~\ref{Fig:4}(c)] vanish. For $-U\le \alpha_1 V_1\le U$, the working point lies at the boundary between the CSRs $(1,1,0)$ and $(1,0,1)$; in fact, the extremes of such interval correspond to triple points, where these two configurations are also degenerate with either one or the other of the two above. The diagonal elements of the QC matrix, corresponding to voltage oscillations along the coordinate directions in the voltage space, display qualitatively different behaviors. In particular, $C_{1;11}$ (green curve) displays two pronounced minima at the triple points, while $C_{1;22}$ and $C_{1;33}$ (red and blue curves) do not. In the region between the triple points, $C_{1;11}$ displays a reduced modulus, while $C_{1;22}$ and $C_{1;33}$ take relatively large negative values. These latter aspects are in qualitative agreement with the classical picture, according to which the voltage oscillations should ideally take place along the direction that is orthogonal to the boundary plane, $\boldsymbol{v}=\boldsymbol{B}_2$, which has components along the $V_2$ and $V_3$ axes, but not along $V_1$. The most negative eigenvalue of the QC matrix, i.e. that with the largest modulus, displays relative minima close to the triple points and an absolute minimum in the intermediate region [Fig.~\ref{Fig:4}(c)], where $C_{1;1}$ is approximately twice the QC obtained for voltage fluctuations along the coordinate axes $V_2$ and $V_3$. The eigenstate corresponding to such eigenvalue, which defines the optimal direction for the voltage oscillation, is characterized by a highly nontrivial dependence on $V_1$, with drastic changes at the triple points [Fig.~\ref{Fig:4}(e)].

In the case of $N=3$, the working-point axis lies in the CSRs $(1,1,1)$ for $-U < \alpha_1 V_1 < U$, on the boundary between $(2,1,0)$ and $(2,0,1)$ for $\alpha_1 V_1<-U$, and between $(0,2,1)$ and $(0,1,2)$ for $\alpha_1 V_1 > U$ (Fig.~\ref{Fig:2}). The extremes of the above intervals correspond to two triple points, where three configurations are degenerate. The behavior of all the plotted quantities is qualitatively different from that of the $N=2$ case, but also presents some clear analogies. In particular, the diagonal elements [Fig.~\ref{Fig:4}(b)] as well as the eigenvalues [Fig.~\ref{Fig:4}(d)] of the QC matrix vanish when the working point is inside the CSR. As to the diagonal elements of the QC matrix, $C_{1;11}$ (green curve) displays two pronounced minima near the triple points and vanishes elsewhere, while $C_{1;22}$ and $C_{1;33}$ (red and blue curves) saturate at their minimal value for $|\alpha_1 V_1| > U$ and vanish in the region between the triple points. The most negative eigenvalue of the QC matrix displays relative minima close to the triple points and a saturation at the absolute minimum value in the external regions, where $C_{1;1}$ is approximately twice the QC obtained for voltage fluctuations along the coordinate axes $V_2$ and $V_3$. The optimal direction for the voltage oscillation is characterized by a highly nontrivial dependence on $V_1$, with abrupt changes at the triple points [Fig.~\ref{Fig:4}(f)].

Clearly, the quantum-mechanical calculations reveal crucial information --- both qualitative and quantitative --- about the QC that could not be predicted from the semiclassical picture, including: the optimal directions of the oscillations in voltage space; the positions and values of the extremal points for the QC along any direction, including the optimal one; the behavior of the QC at triple points in the CSR graph, where three boundary (hyper)planes intersect; the existence of plateaus in the dependence of the QC on the position of the working point. Therefore, the semiclassical considerations are useful to preselect the relevant regions in voltage space, where interesting features of the QC occur, but accurate indications can only be obtained from the quantum-mechanical treatment.

\paragraph{Quantitative integration of the semiclassical picture.}
A peak in the QC is characterized not only by its position in the voltage space, but also by its height and width. These features can be related to those which characterize the avoided level crossing that the Hamiltonian ground and first-excited states undergo as $\boldsymbol{V}$ crosses the boundary hyperplane separating two CSRs with occupations $\boldsymbol{n}$ and $\boldsymbol{n}'$. Besides, if other hyperplanes are sufficiently far away in the voltage space, then the transition between the stability regions can be approximately described within a reduced subspace. This is spanned by different quantum states, sharing the dot occupations $\boldsymbol{n}$ or $\boldsymbol{n}'$, and differing from one another in terms of spin orientations. 
 
In the simplest case, the avoided level crossing involves only two basis states, one for each CSR. The effective Hamiltonian $H^{\rm eff}$ on this two-dimensional space has the general form
\begin{align}
H^{\rm eff} = \left[ \begin{matrix} E_{\boldsymbol{n}}(\boldsymbol{V}) & \tau  \\ \tau  & E_{\boldsymbol{n}'}(\boldsymbol{V}) \end{matrix} \right] \,,
\label{Heff}
\end{align}
where $\tau$ is the transition amplitude between the two basis states, due to the terms of the Hamiltonian that do not conserve the dot occupations, and generally depends on $\boldsymbol{n}$, $\boldsymbol{n}'$ and $\boldsymbol{V}$. In the case of the Hubbard model, $\tau$ is related to the hopping processes, which can be of different orders. The eigenvalues of $H_{\rm eff}$ read:
\begin{align}
E^{\rm eff}_{\pm} = \frac{1}{2} \left[ E_{\boldsymbol{n}}(\boldsymbol{V}) + E_{\boldsymbol{n}'}(\boldsymbol{V})\pm \sqrt{\delta^2 (\boldsymbol{V}) + 4 |\tau|^2} \right] \,,
\end{align}
where $\delta(\boldsymbol{V}) = E_{\boldsymbol{n}}(\boldsymbol{V}) - E_{\boldsymbol{n}'}(\boldsymbol{V})$. Under the assumptions that the diagonal and off-diagonal elements of $H^{\rm eff}$ are respectively linearly dependent on $\boldsymbol{V}$ and independent of $\boldsymbol{V}$, the elements of the ground-state QC matrix at the avoided crossing [$\delta(\boldsymbol{V}) = 0$] are
\begin{align}
\left. \frac{\partial^2 E^{\rm eff}_{-}}{\partial V_i \partial V_j} \right|_{\delta(\boldsymbol{V}) = 0} = - \frac{1}{4 \left| \tau \right|} \frac{\partial \delta(\boldsymbol{V}) }{\partial V_i}  \frac{\partial \delta(\boldsymbol{V}) }{\partial V_j} \,.
\end{align} 
This indicates that the minimum value of the QC along any direction is inversely proportional to $|\tau|$. If the two basis states are coupled by a first-order hopping process, then the transition amplitude $\tau$ is proportional to the hopping parameter $T$. This is the case, e.g., for the transition between the CSRs $(1,1,1)$ and $(0,2,1)$, taking place as $\boldsymbol{V}$ moves along the line $\boldsymbol{B}_1$ (Fig.~\ref{Fig:3}). In this case, one can in fact solve the reduced model analytically and show that $\tau = \sqrt{2} T$ (Appendix \ref{app:analytical1}).

\begin{figure}
\centering
\includegraphics[scale=0.5]{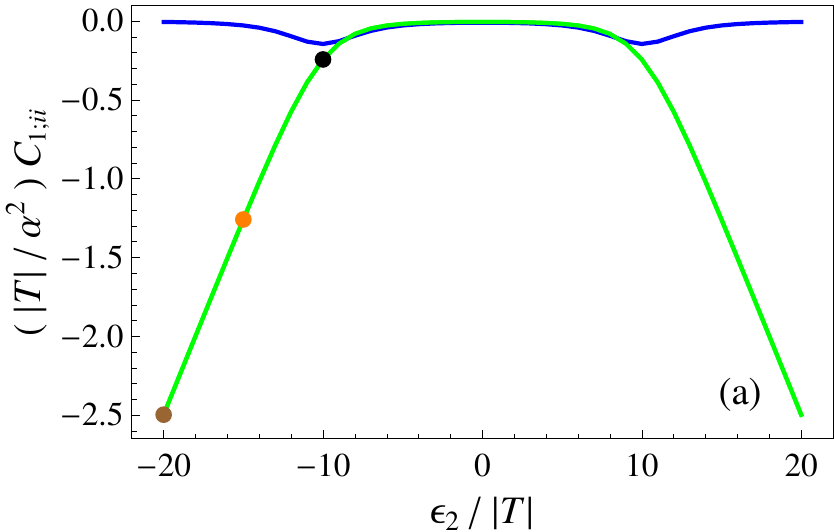} 
\includegraphics[scale=0.5]{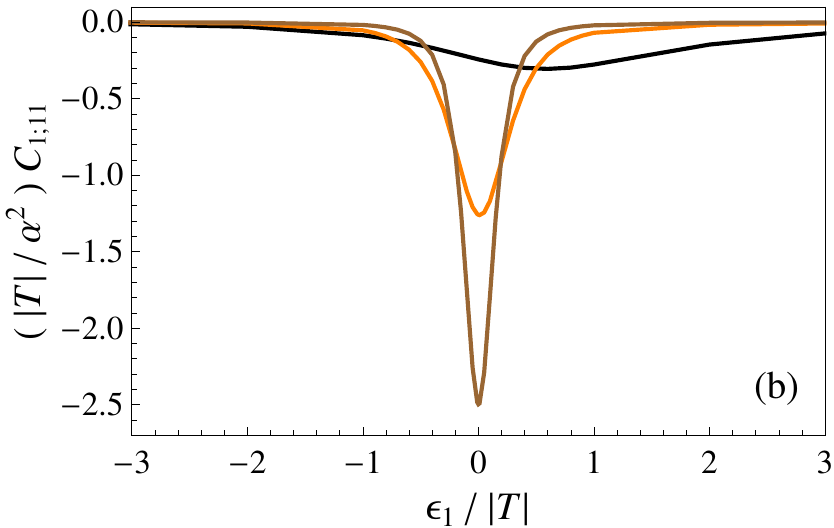} 
\includegraphics[scale=0.5]{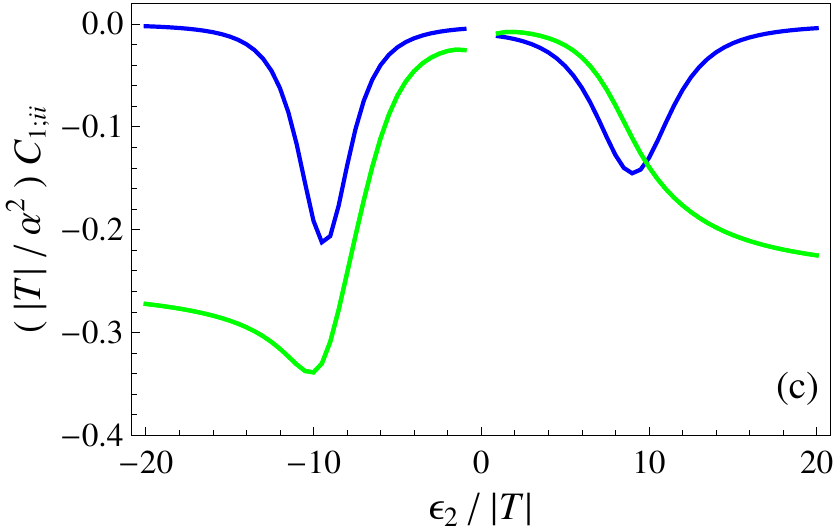} 
\includegraphics[scale=0.5]{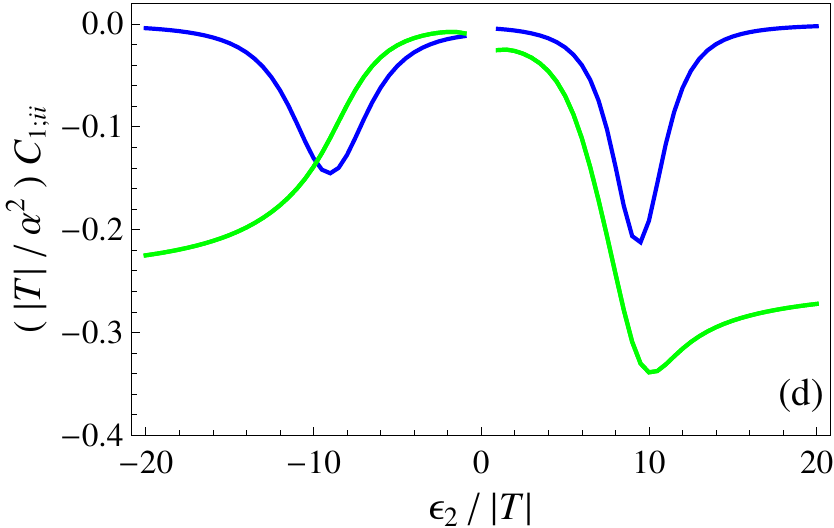} 
\caption{(a) Diagonal elements of the ground-state QC matrix $C_{1;ii}$ for the open chain with $M = 3$ sites and $N = 3$ electrons, as functions of the central onsite energy $\epsilon_2$, for $\epsilon_1 = \epsilon_3 = 0$. Colors correspond to the value of $i$: green $\rightarrow i = 1$ and $i = 3$ (the curves coincide), blue $\rightarrow i = 2$. The three dots correspond to the values of $\epsilon_2$ considered in panel (b). In panel (b) we plot $C_{1; 11}$ at the working point $(\epsilon_1, \epsilon_2, 0)$ as a function of $\epsilon_1$, for three selected values of $\epsilon_2$: black $\rightarrow \epsilon_2 = -10 |T|$, orange $\rightarrow \epsilon_2 = -15 |T|$, brown $\rightarrow \epsilon_2 = -20 |T|$. Calculations are done for $U = 10 |T|$, and $\alpha_i \equiv \alpha$ $\forall i$. Panels (c) and (d): Diagonal elements of the ground-state QC matrix in the same domain and with the same conventions as in panel (a), but for the case of the closed ring, respectively for $T<0$ and $T>0$.  }
\label{Fig:5}
\end{figure}

However, there are also pairs of CSRs that are coupled by higher-order hopping processes, whose amplitude depends not only on the hopping parameter $T$, but also on the applied voltages. As a consequence, also the height and width of the QC peaks can be tuned by varying the applied voltages. This is the case, for example, of the open chain when $M=N=3$ and $\boldsymbol{V}_0$ lies on the $V_2$ axis [vertical axis of Fig.~\ref{Fig:2}(b)]. The obtained behavior of $C_{1;22}$ is qualitatively similar to that observed in previous examples, with two peaks in the vicinity of the triple points and vanishing values elsewhere [Fig.~\ref{Fig:5}(a)]. More interestingly, the diagonal elements $C_{1;11}$ and $C_{1;33}$, identical by symmetry if $\alpha_i \equiv \alpha$ is independent of $i$, display no absolute minimum and decrease monotonically for increasing values of $|\epsilon_2|$. This peculiar behavior can be explained and reproduced within a reduced model including four basis states (Appendix \ref{app:effective model}). The analytical solutions of the model, which are in good agreement with the numerical solution of the full Hamiltonian, show in fact that 
\begin{align}
C^{\rm eff}_{1;11} = C^{\rm eff}_{1;33}  \approx - \alpha^2\frac{|\epsilon_2|}{4 T^2}   \label{Cii eff}
\end{align}
for $|\epsilon_2| \gg U, |T|$. Unlike the previously considered case, the transition between the configurations $(2,0,1)$ and $(1,0,2)$ involves a second-order hopping process, with an interconnecting configuration $(1,1,1)$ whose energy difference with respect to the other two depends on $V_2$. As a result, the effective amplitude of the transition, as results from Eq.~\eqref{Cii eff}, is proportional to $T^2$ and inversely proportional to $|V_2|$. Increasing values of $|V_2|$ beyond $U$ thus give rise to peaks of increasing height; correspondingly, the width of the curve along the orthogonal axis $V_1$ decreases [Fig.~\ref{Fig:5}(b)]. The possibility of tuning the shape of the QC peak by modifying the working point potentially has important practical implications. In fact, a higher peak in principle enhances the signal and thus favors the discrimination of the states based on the contrast in terms of QC. On the other hand, an increased peak height also comes with a reduced width of the avoided level crossing between the relevant quantum states. This typically implies more stringent requirements in terms of the frequency of the probe, if one wants to preserve the adiabatic character of the dynamics. Besides, a narrow peak also implies a tighter upper bound on the amplitude of the probe, in order to avoid overdrive errors \cite{Derakhshan20a}.

As mentioned above, the effect depicted in Fig.~\ref{Fig:5}(a) and (b) is due to transitions between configurations which, in the case of an open chain, are connected by second-order hopping processes. In the case of the closed ring, instead, the configurations $(2,0,1)$ and $(1,0,2)$ [as well as $(1,2,0)$ and $(0,2,1)$] are connected by a first-order hopping process. As a result, the diagonal elements of the QC matrix are different from those of the open chain, and they are given by the curves depicted in Fig.~\ref{Fig:5}(c) and (d), respectively for $T<0$ and $T>0$. The main differences are that: (i) $C_{1; 11} = C_{1; 33}$ does not diverge for $\left| \epsilon_2 \right| \rightarrow \infty$; (ii) energies (and, therefore, QCs) are not even functions of $T$; (iii) in the considered domain, at $\epsilon_2 = 0$ the closed ring has a degeneracy between two doublet states, due to a local symmetry of the Hamiltonian. Therefore, close to that point the definition of the QC as the Hessian matrix of an eigenenergy loses its meaning and should be replaced with the Hessian matrix of an appropriately defined mixed state.

\section{Conclusions and outlook}
\label{sec: conclusions}

In conclusion, we have provided a multidimensional generalization of the QC for a generic system of interacting particles in the presence of multiple metal gates. This leads to the introduction of the QC matrix, whose diagonalization allows one to identify the optimal direction in the voltage space along which to induce oscillations in order to maximize and detect the QC. As a possible application of this approach, we have considered a Hubbard model describing a linear quantum-dot array. Based on the expressions obtained within such model, the potentially relevant working points are identified with the hyperplanes defining the boundaries between charge stability regions, which are determined by means of a semiclassical picture. On the one hand, the quantum approach confirms the relevance of such boundaries, where the QC displays its maximal absolute values, and shows that the optimal direction in general coincides with the normal to the hyperplane. On the other hand, it accounts for a number of additional quantitative and qualitative features, including the dependence of the QC on the process responsible for the coupling between charge configurations and its behavior at the intersections between different hyperplanes. Altogether, these results clearly shows the possibility of using the QC as a means to discriminate between different spin and charge states, along the lines of what has been done in the readout of single- and two-spin (singlet-triplet) qubits. 

The presented results can be extended in different respects. In particular, some of the assumptions that have been introduced in the Hubbard model with the aim of simplifying the discussion can be easily removed without significantly modifying the approach. If, for example, the assumption concerning the local character of the coupling between quantum dots and metal gates [Eq.~\eqref{eq:linearity}] is not valid, then the expressions derived for the Hubbard model no longer apply to the gate voltages, but they do apply to the virtual gates. In the presence of a nonlinear dependence of the onsite energy on the applied voltage, an additional term in the expression of the QC would result from the differentiation of Eq.~\eqref{Hellmann-Feynman Hubbard}. Also, the separations between charge stability regions would no longer be given by hyperplanes, but rather by more general surfaces. This would require a revisiting of the semiclassical approach, but the general equations derived in the paper would still apply. We also note that none of the general equations concerning the QC matrix depend on the linear gate arrangements considered in the presented applications. Different geometries would be simply characterized by different patterns of interdot couplings, and thus by different eigenstates and eigenenergies. We have shown how this can affect the measurable QCs by comparing the two exemplary cases of an open chain and a closed ring of Hubbard quantum dots. In the case of a more general Hubbard model, including more than one orbital per site, the above equations can be easily generalized by replacing the dot occupations with those of the individual orbitals. Finally, the dependence of also other Hamiltonian parameters --- such as the interdot tunneling --- on the voltages might in principle be included, and could in practice become relevant in a regime where small changes in the voltage can induce transitions between configurations with different values of combined expectation values, including both dot occupations and interdot coherences \cite{e25010082}. 

Beyond the Hubbard model, the equations reported in Section \ref{sec: MQC} can be applied to a generic few-particle state affected by a combination of gate voltages. In particular, combining the present formalism with a detailed modeling of spin qubits in single and coupled quantum dots \cite{PhysRevB.98.155319,Bellentani2021a,Secchi2021a,PhysRevB.104.235302,PhysRevLett.129.066801,PhysRevB.104.195421} would allow realistic theoretical characterizations of the quantum capacitance and possibly provide criteria for its enhancement through the tuning of the device parameters. This might in turn allow an improvement in the readout fidelity of different spin-based qubits. 

\acknowledgments

The authors acknowledge financial support from IQubits (Call: H2020--FETOPEN--2018--2019--2020--01, Project ID: 829005) and from the PNRR MUR project PE0000023-NQSTI, and fruitful discussions with Andrea Bertoni, Michele Spasaro, Domenico Zito, and Alessandro Crippa.

\appendix

\section{Symmetry of the QC matrix}
\label{app:symmetry proof}

We now demonstrate the symmetry of the QC matrix. The first term in the right-hand side of Eq.~\eqref{QC after HF multigate} is evidently symmetric under exchange of indexes $i$ and $j$. We need to address the second term,
\begin{align}
  \int d \boldsymbol{r}     \frac{\partial W(\boldsymbol{r},
\boldsymbol{V})}{\partial V_i}   \frac{ \partial n_k(\boldsymbol{r})
}{\partial V_j}    \,.
  \label{of interest}
\end{align}
We start by writing
\begin{align}
\frac{\partial n_k(\boldsymbol{r})}{\partial V_j}   = \frac{\partial \big< \psi_k \big|}{\partial V_j} 
\hat{n}(\boldsymbol{r}) \big| \psi_k \big> +  \big< \psi_k \big|
\hat{n}(\boldsymbol{r}) \frac{\partial \big| \psi_k \big>}{\partial V_j}
  \,,
\label{exp 1}
\end{align}
where $\hat{n}(\boldsymbol{r})$ is the (many-body) density operator. If either $\big| \psi_k \big>$ is non-degenerate, or its degeneracy is due to a global symmetry of the Hamiltonian that holds independently of $\boldsymbol{V}$ (e.g., time-reversal), then the derivative of the state vector $\big| \psi_k \big>$ with respect to parameter $V_j$ can be expressed as 
\begin{align}
\frac{\partial \big| \psi_k \big>}{\partial V_j} & = \left( \big< \psi_k \big| \frac{\partial \big| \psi_k \big> }{\partial V_j} \right) \big| \psi_k \big>      \nonumber \\
& \quad  + \sum_{m : E_m \neq E_k}
\frac{1}{E_k - E_m} \big< \psi_m \big| \frac{\partial \hat{H}}{\partial
V_j} \big| \psi_k \big> \big| \psi_m \big> \,,
\label{exp 2}
\end{align}
where $\hat{H}$ is the Hamiltonian, with eigenstates $\big| \psi_m \big> $ and eigenenergies $E_m$. The first term in the right-hand side of Eq.~\eqref{exp 2} is purely imaginary due to the normalization condition $\big< \psi_k \big| \psi_k \big> = 1$, and will be irrelevant in what follows. Equation \eqref{exp 2} follows from the so-called out-of-diagonal Hellmann-Feynman theorem \cite{Singh89}.

Since the Hamiltonian depends on $V_j$ only via the term $\hat{W}$ [see Eq.~\eqref{general H}], the derivative of the Hamiltonian reads
\begin{align}
\frac{\partial \hat{H}}{\partial V_j} = \int d \boldsymbol{r}' \,
\frac{\partial W(\boldsymbol{r}', \boldsymbol{V})}{\partial V_j} \,
\hat{n}(\boldsymbol{r}') \,.
\label{exp 3}
\end{align}
Combining Eqs.~\eqref{of interest}-\eqref{exp 2} and \eqref{exp 3},
one obtains
\begin{align}
& \int d \boldsymbol{r}     \frac{\partial W(\boldsymbol{r},
\boldsymbol{V})}{\partial V_i}   \frac{ \partial n_k(\boldsymbol{r})
}{\partial V_j}    \nonumber \\
& =   \int d \boldsymbol{r}   \int d \boldsymbol{r}' \,   \frac{\partial
W(\boldsymbol{r}, \boldsymbol{V})}{\partial V_i}
\frac{\partial W(\boldsymbol{r}', \boldsymbol{V})}{\partial V_j}
\nonumber \\
& \quad \times \sum_{m : E_m \neq E_k} \frac{n_{km}(\boldsymbol{r})  \, 
  n_{mk}(\boldsymbol{r}') + n_{km}(\boldsymbol{r}')  \,  n_{mk}(\boldsymbol{r})    }{E_k(\boldsymbol{V}) - E_m(\boldsymbol{V}) }   \,,
  \label{pathological term}
\end{align}
where $n_{km}(\boldsymbol{r}) \equiv \big< \psi_k \big| \hat{n}(\boldsymbol{r}) \big| \psi_m \big>$, and the purely imaginary first term in the right-hand side of Eq.~\eqref{exp 2} cancelled out in the sum with its complex conjugate.

It is then seen that, if there are points in the parameter space $\boldsymbol{V} = \boldsymbol{V}_{m k}$ such that $ E_m(\boldsymbol{V}_{m k}) = E_k(\boldsymbol{V}_{m k})$ for a certain $m$, at those points a denominator in Eq.~\eqref{pathological term} vanishes, and consequently the QC (or some of its elements) might not exist. The actual (non)-existence depends on the limiting behavior of the whole expression \eqref{pathological term} when $\boldsymbol{V} \rightarrow \boldsymbol{V}_{m k}$, and should be evaluated case by case.

Physically, the working points at which an energy degeneracy occurs are precisely the points where the adiabatic theorem does not apply, and the Hessian matrix of an eigenenergy is not defined. Consistently, at those points the definition of the QC matrix should be based on the expectation value of the energy, ${\rm Tr}\left( \hat{H} \hat{\rho} \right)$, rather than on a specific eigenenergy.  

Away from the degeneracy points, the QC is well-defined and Eq.~\eqref{pathological term} is clearly symmetric under the exchange $i \leftrightarrow j$, which completes the proof that the whole QC matrix \eqref{QC after HF multigate} is symmetric.

\section{Hyperplanes separating the CSRs}
\label{app: hyperplanes}

The equation for the hyperplane separating the two CSRs specified, respectively, by occupations $\boldsymbol{n}$ and $\boldsymbol{n}'$ is given by $E_{\boldsymbol{n}}(\boldsymbol{V}) = E_{\boldsymbol{n}'}(\boldsymbol{V})$. For the Hubbard model [Eq.~\eqref{classical energy}], the hyperplane equations have a simple form in the $\boldsymbol{\epsilon}$ space, where $\epsilon_i = \alpha_i V_i$:
\begin{align}
\boldsymbol{\epsilon} \cdot \left( \boldsymbol{n} - \boldsymbol{n}' \right) + \frac{U}{2} \left( \left| \boldsymbol{n} \right|^2 -\left| \boldsymbol{n}' \right|^2 \right) = 0 \,.
\label{hyperplane equation}
\end{align}
Here we have used of the fact that the CSRs correspond to the same total number of particles: $\sum_{i=1}^M n_i = \sum_{i=1}^M n'_i = N$. The normal $\boldsymbol{B}'$ to the hyperplane that represents the boundary between two CSRs in the $\boldsymbol{\epsilon}$ space thus has components given by the difference in the occupations of the $M$ dots. The corresponding hyperplane in the voltage space is defined by a normal vector $\boldsymbol{B}$ whose components coincide with those of $\boldsymbol{B}'$, up to multiplicative factors $1/\alpha_i$. While these equations apply to any pair of CSRs, the boundaries that are considered in the present paper are specifically the ones between ground-state configurations. In these cases, $\boldsymbol{n}$ and $\boldsymbol{n}'$ turn out to differ by two components $i$ and $j$, with $n_i-n_i'=n_j'-n_j=1$.

Using again the particle-number conservation, it is easily shown that the vector $\boldsymbol{B}'$ is also perpendicular to the $M$-dimensional vector $(1, 1, \ldots, 1)$ in the $\boldsymbol{\epsilon}$ space. Therefore, the vector $\boldsymbol{D} = ( \alpha_1^{-1}, \ldots, \alpha_M^{-1} )$ in $\boldsymbol{V}$ space is perpendicular to any $\boldsymbol{B}$ that defines a boundary hyperplane, and thus parallel to all the hyperplanes separating adjacent CSRs, as stated in the main text.

\section{Analytical model for the QC related \\ to the $(1,1,1) \rightarrow (0,2,1)$ transition}
\label{app:analytical1}

We here consider the Hubbard model in the case of $M=3$ and $N=3$, and reduce it to an approximate, analytically solvable form in order to effectively describe the transition between the CSRs $(1,1,1)$ and $(0,2,1)$ along the direction $\boldsymbol{\epsilon} = \epsilon_1 (1, -1, 0)$ with $\epsilon_1 > 0$ (Fig.~\ref{Fig:3}). 

Since ${\bf S}^2$ and $S_z$ commute with the Hamiltonian, we can restrict our theory to a subspace with a defined, ground-state value of both observables, namely $S=S_z = 1/2$ (the same analysis applies to the subspace $S_z = -1/2$). The basis states involved in the $(1,1,1) \leftrightarrow (0,2,1)$ transition are: the set $\{ | \uparrow, \uparrow, \downarrow \rangle$, $| \uparrow, \downarrow, \uparrow \rangle$, $| \downarrow, \uparrow, \uparrow \rangle \}$, with energy $\epsilon_1 + \epsilon_2 + \epsilon_3$, and the state $\big| 0, \uparrow \downarrow, \uparrow \big>$ with energy $U  + 2 \epsilon_2 + \epsilon_3$ [compare with Eq.~\eqref{classical energy}]. On the domain $\boldsymbol{\epsilon} = \epsilon_1 (1, -1, 0)$, the latter becomes the classical ground state for $\epsilon_1 > U/2$, where the transition between the two CSRs occurs. Other states have higher energies and will be neglected in order to  obtain an analytical solution.

Within the selected 4-dimensional subspace, state $\left. \big|  \uparrow, \uparrow, \downarrow \big> \right.$ and the combination $\frac{1}{\sqrt{2}} \left( \big| \uparrow, \downarrow, \uparrow \big> + \big| \downarrow, \uparrow, \uparrow  \big>   \right)$ are eigenstates of the Hamiltonian with energy $\epsilon_1 + \epsilon_2 + \epsilon_3$, and do not couple with the other states. The other two states, namely $\frac{1}{\sqrt{2}} \left( \big| \uparrow, \downarrow, \uparrow \big> -   \big| \downarrow , \uparrow, \uparrow \big>  \right)$ and $\big| 0, \uparrow \downarrow, \uparrow \big>$, are coupled by tunneling, and determine the ground state. In particular, the relevant tunneling process is that between site $1$ and site $2$, which is the same in the two cases considered here (open chain and closed ring). Therefore, in both cases, the original system is effectively reduced to a two-state one, whose Hamiltonian reads:
\begin{align}
H^{\rm eff} = \left( \begin{matrix} \epsilon_1 + \epsilon_2 + \epsilon_3 &  \sqrt{2} T \\ 
\sqrt{2} T &   U  + 2 \epsilon_2 + \epsilon_3 \end{matrix} \right) \,.
\end{align}
This has the same form as Eq.~\eqref{Heff}, with the correspondences $E_{\boldsymbol{n}}(\boldsymbol{V}) \leftrightarrow \epsilon_1 + \epsilon_2 + \epsilon_3$, $E_{\boldsymbol{n}'}(\boldsymbol{V}) \leftrightarrow U  + 2 \epsilon_2 + \epsilon_3$, and $\tau \leftrightarrow \sqrt{2} T$. 

The ground-state energy is given by
\begin{align}
E^{\rm eff}_1  = \frac{1}{2} \! \left[ \epsilon_1  + 3 \epsilon_2 + 2 \epsilon_3  + U   - \sqrt{8 T^2 + \left( \epsilon_2 - \epsilon_1 + U \right)^2    }   \right] .
\end{align}
From this expression, one can derive the QC matrix, which has two zero eigenvalues, corresponding to the eigenvectors $(0,0,1)$ and $(1,1,0)$, and a negative-defined eigenvalue given by
\begin{align}
C^{\rm eff}_{1; 1} = - \alpha^2 \frac{8 T^2}{\left[ 8 T^2 + \left( \epsilon_2 - \epsilon_1 + U \right)^2 \right]^{3/2}} \,,
\label{C min (0,-1,1)}
\end{align}
corresponding to the normalized eigenvector $\boldsymbol{v}_{1;1} = 2^{-1/2} (1, -1, 0)$. Therefore, within this reduced model, the eigenvectors are independent of $\boldsymbol{\epsilon}$, and the eigenvalue with the maximum modulus coincides with the QC produced by a perturbation $\delta \boldsymbol{V}$ perpendicular to the hyperplane separating the two CSRs, in agreement with the numerical results shown in Fig.~\ref{Fig:3}. 

The minimum value of $C^{\rm eff}_{1; 1}$ [Eq.~\eqref{C min (0,-1,1)}] on the domain $\boldsymbol{\epsilon} = \epsilon_1 (1,-1,0)$ is reached at $\epsilon_1 = U/2$, and is given by
\begin{align}
\min\left\{C^{\rm eff}_{1; 1} \right\} = - \alpha^2 \frac{1}{2 \sqrt{2} \left|   T  \right| } \approx - 0.3536 \frac{\alpha^2}{|T|} \,,
\end{align}
in good agreement with the numerical results obtained for both the open chain and the closed ring regarding the eigenvalue,
\begin{align}
& \min\left\{C_{1; 1} \right\}   \approx - 0.3621 \frac{\alpha^2}{|T|} \quad {\rm (chain)}\,,  \nonumber \\
& \min\left\{C_{1; 1} \right\}   \approx - 0.3303 \frac{\alpha^2}{|T|} \quad {\rm (ring)}\,,  \nonumber \\
\end{align}
as well as the QC along the classically-suggested optimal direction,
\begin{align}
& \min\left\{C_{1; \frac{1}{\sqrt{2}} (1,-1,0)} \right\}   \approx - 0.3616 \frac{\alpha^2}{|T|} \quad {\rm (chain)} \,, \nonumber \\
& \min\left\{C_{1; \frac{1}{\sqrt{2}} (1,-1,0)} \right\}   \approx - 0.3302 \frac{\alpha^2}{|T|} \quad {\rm (ring)} \,. 
\end{align}
The small differences between the numerical results for the chain and the ring (shown in Fig.~\ref{Fig:3}), as well as the small deviation from zero of one of the eigenvalues of the QC matrix, are due to higher-order tunneling processes that are not included in the reduced analytical model discussed in this Appendix.

\section{Analytical model for the QC \\ at the boundary between charge stability \\ regions $(1,2,0)$ and $(0,2,1)$, for an open chain}
\label{app:effective model}

We now formulate an analytical model, analogous to the one discussed in Appendix \ref{app:analytical1}, for the transitions along the line $\epsilon_1=\epsilon_3=0$, with $M=N=3$ and $S_z = 1/2$, in the case of an open chain. In particular, we derive the expressions of the diagonal elements of the QC matrix, in order to account for the peculiarities obtained numerically [Fig.~\ref{Fig:5}(a)]. 

The aim is to achieve an accurate description for the range of values $\epsilon_2 \lesssim -U$, where the set of lowest-energy configurations include: the three $(1,1,1)$ configurations 
$\left\{ \big| \uparrow, \uparrow, \downarrow \big> , \big| \uparrow, \downarrow, \uparrow \big> , \big| \downarrow, \uparrow, \uparrow \big>\right\}$, 
with energy $\epsilon_1 + \epsilon_2 + \epsilon_3$; the $(1,2,0)$ configuration 
$\big| \uparrow, \uparrow \downarrow, 0 \big>$, 
with energy $U + \epsilon_1 + 2 \epsilon_2$; 
the $(0,2,1)$ configuration $\big| 0, \uparrow \downarrow, \uparrow \big>$, with energy $U + 2 \epsilon_2 + \epsilon_3$. 
The Hubbard Hamiltonian is projected on this five-dimensional subspace, ignoring the couplings with higher-energy states. The $(S = 3/2)$-state $\frac{1}{\sqrt{3}} \big(  \left. \big| \uparrow, \uparrow, \downarrow \big> \right. + \big| \uparrow, \downarrow, \uparrow \big> + \big| \downarrow, \uparrow, \uparrow \big> \big)$ is an eigenstate with energy $\epsilon_1 + \epsilon_2 + \epsilon_3$, and does not couple with the other states, which thus define an independent four-dimensional subspace.

The analytical diagonalization of the Hamiltonian is not possible for arbitrary values of the parameters. It is possible, however, on the line $\epsilon_1 = \epsilon_3 = 0$, where the ground-state energy reads
\begin{align}
E^{\rm eff}_{1} = \frac{1}{2} \left[ 3 \epsilon_2 + U -
\sqrt{\left( \epsilon_2 + U \right)^2 + 12 T^2  } \right] \,.
\label{GS red}
\end{align}
This energy has an avoided crossing with that of the fourth excited state ($E^{\rm eff}_{5}$) in the five-dimensional reduced space; analogously, there is an avoided crossing between the first and third excited energies. The avoided crossing between $E^{\rm eff}_{1}$ and $E^{\rm eff}_{5}$ can be described in terms of a two-state reduced model with $\tau = \sqrt{3} T$, because the transition along the line $\epsilon_1 = \epsilon_3 = 0$ requires a first-order tunneling process. However, such picture cannot be applied to the transverse transitions, which require the full four-dimensional subspace. 

From Eq.~\eqref{GS red}, by differentiating twice with respect to $V_2$, one can directly obtain the following diagonal element of the QC matrix:
\begin{align}
C^{\rm eff}_{1; 2 2} = - \alpha_2^2 \frac{6 T^2}{\left[ \left(
\epsilon_2 + U \right)^2 + 12 T^2 \right]^{3/2}} \,.
\label{GS red QC 22}
\end{align}
The derivation of the two other diagonal elements is less
straightforward. It can be seen that, by symmetry, $C^{\rm eff}_{1; 1 1 } / \alpha_1^2 = C^{\rm eff}_{1; 3 3} / \alpha_3^2$. In order to determine $C^{\rm eff}_{1; 1 1 }$, one would need, in principle, to diagonalize the reduced Hamiltonian in the case of $\epsilon_3 = 0$ and $\epsilon_1, \epsilon_2 \neq 0$, find the ground-state energy, and take its second derivative with respect to $\epsilon_1$. However, if both $\epsilon_1$ and $\epsilon_2$ are $\neq 0$, the ground-state energy is determined by a 4th-degree polynomial equation that cannot be solved analytically. Still, with some manipulations on that equation, and the knowledge of Eq.~\eqref{GS red}, it is possible to derive the analytical expression for $C^{\rm eff}_{1; 1 1 }$, even without having an explicit expression for the ground-state energy. 

The expression we finally obtain for the two diagonal elements of the QC matrix with $i \in \lbrace 1,3 \rbrace$ reads: 
\begin{align}
C^{\rm eff}_{1; ii}   = - \frac{\alpha_i^2}{8 T^2} \left\{
\frac{\left[ (\epsilon_2 + U)^2 + 9 T^2 \right]^2 + 3 T^4 }{ \left[ \left( \epsilon_2 + U \right)^2
+ 12 T^2  \right]^{3/2}}   - (\epsilon_2 + U)   \right\} .
\label{GS red QC 11}   
\end{align}
Equations \eqref{GS red QC 11} and \eqref{GS red QC 22} are in good agreement with the results obtained from the numerical solution of the Hubbard model. In particular, the analytical model accounts for the large-$|\epsilon_2|$ behavior of $C_{1; 11}$ and $C_{1; 33}$. In fact, when $|\epsilon_2|   \gg U$ and $|\epsilon_2|   \gg |T|$, one finds the limiting behavior given by Eq.~\eqref{Cii eff}, which accounts for the indefinite increase of $|C_{1; ii}|$ with $|\epsilon_2|$ [Fig.~\ref{Fig:5}(a)].

\end{document}